\documentclass[seceq]{ptptex}

\usepackage{graphicx}

\title{Anomalous Transport Processes in Anisotropically Expanding 
       Quark-Gluon Plasmas}

\author{Masayuki \textsc{Asakawa}$^1$, Steffen A. \textsc{Bass}$^2$
        and Berndt \textsc{M\"uller}$^{2,3}$}

\inst{$^1$Department of Physics, Osaka University, Toyonaka 560-0043, Japan \\
      $^2$Department of Physics, Duke University, Durham, NC 27708, USA \\
      $^3$Yukawa Institute of Theoretical Physics, Kyoto University,
             Kyoto 606-8502, Japan}

\abst{We derive an expression for the anomalous viscosity in an 
anisotropically expanding quark-gluon-plasma, which arises from
interactions of thermal partons with dynamically generated 
color fields. The anomalous viscosity dominates over the 
collisional viscosity for large velocity gradients or weak coupling.
This effect may provide an explanation for the apparent ``nearly
perfect'' liquidity of the matter produced in nuclear collisions
at RHIC without the assumption that it is a strongly coupled state.}

\begin{document}

\maketitle

\section{Introduction}

Measurements of the so-called elliptic flow parameter $v_2$ 
of hadrons emitted in noncentral collisions of heavy nuclei at 
the Relativistic Heavy Ion Collider (RHIC)  are in remarkably 
good agreement with the predictions of ideal relativistic fluid 
dynamics.\cite{RHIC_WP} \ In order to achieve the agreement with 
the data, the hydrodynamical calculations need to assume a rapid 
initial equilibration within a time \cite{Heinz:2001xi} 
$\tau_{\rm i} < 1$ fm/c and an extremely small ratio of the shear 
viscosity $\eta$ of the fluid to its entropy density 
$s$.\cite{Teaney:2003pb} \ In fact, the ratio $\eta/s$ cannot be much
larger than the conjectured lower bound $(4\pi)^{-1}$ for this 
quantity.\cite{Kovtun:2004de} 

These findings have led to the 
general conclusion that the matter produced at RHIC is strongly
interacting with a large opacity, i.e. small mean free path.
However, this does not necessarily have to be the case:
In a previous Letter \cite{Asakawa:2006tc} we have argued that the 
unavoidable color field instabilities in a rapidly expanding quark-gluon
plasma generate an anomalous viscosity, which is smaller than 
the collisional viscosity in the weak coupling limit and for early times
after the onset of the longitudinal expansion.
The expression derived in Ref.~\citen{Asakawa:2006tc} 
for the anomalous viscosity,
\begin{equation}
\frac{\eta_{\rm A}}{s} 
  = \frac{1}{g^{3/2}} \left(\frac{(N_c^2-1)T\tau}{10 b_0 N_c}\right)^{1/2} \, ,
\label{eq:eta-AA}
\end{equation}
where $N_c$ denotes the number of colors and $b_0$ is a presently 
unknown numerical coefficient, is parametrically (in the coupling 
constant $g$) smaller than the collisional viscosity 
\cite{Arnold:2000dr}
\begin{equation}
\frac{\eta_{\rm C}}{s} = \frac{d_f}{g^4 \ln(\sqrt{4\pi}/g)} ,
\label{eq:eta-C}
\end{equation}
where $d_f \approx 5$ for three light quark flavors. The 
anomalous viscosity is generated by the diffusive transport of 
quasithermal partons in the turbulent color fields created by
Weibel-type instabilities. Such instabilities always arise 
when the parton momentum distribution is 
anisotropic,\cite{Mrowczynski:1988dz,Mrowczynski:1993qm,Romatschke:2003ms} \ 
as is necessarily the case when the quark-gluon plasma expands with 
a preferred direction. The nonlinear interaction of the unstable modes
has been shown to result in a turbulent cascade of field energy into 
increasingly short wavelength modes,\cite{Arnold:2005ef,Arnold:2005qs} \ 
the characteristic signature of a turbulent plasma.

The purpose of this article is to present additional details
of the calculation outlined in our previous Letter. 
The article is structured as follows. In Section 2 we discuss the
evidence from lattice QCD calculations for a quasi-particle nature
of the quark-gluon plasma, even in the vicinity of the transition 
temperature $T_c$. This evidence is derived from the off-diagonal 
flavor structure of the quark susceptibilities. On the other hand,
the RHIC data on jet quenching combined with entropy constraints 
demand a plasma almost entirely composed of gluons. The anomalous
transport properties owing to the presence of intense coherent color 
fields can alleviate these constraints. In Section 3 we give a 
heuristic derivation of the anomalous viscosity $\eta_{\rm A}$. 
In Section 4 we derive the transport equation for quarks and gluons
in a turbulent plasma. Section 5 contains a brief reminder of the
definition of the shear viscosity. In Section 6  we present details 
of the formal derivation of $\eta_{\rm A}$ within the framework of 
the theory of transport phenomena in turbulent plasmas. 
We summarize our results in Section 7.
Two appendices give an alternative derivation of the transport 
equation in the presence of plasma turbulence and provide details
on the plasma instability structure, respectively.

\section{The Quark-Gluon Plasma at RHIC}

There is now overwhelming consensus that matter in (approximate) 
thermal equilibrium and with an energy density far in excess of 
1 GeV/fm$^3$ is produced in nuclear collisions at 
RHIC.\cite{RHIC_WP,Ludlam:2005gx} \ The evidence leading to this 
conclusion rests on two key observations. Firstly, all stable hadrons 
including multistrange baryons are emitted with chemical 
equilibrium abundances\cite{Rafelski:2001hp,Braun-Munzinger:2001ip} 
and thermal transverse momentum spectra that are boosted by a 
collective transverse (``radial'') flow.\cite{Adams:2003xp,Adcox:2003nr} \
Secondly, the hadron spectra measured in noncentral collisions 
show an azimuthal quadrupole anisotropy (``elliptic'' flow)
characteristic of the fluid dynamical expansion of a fireball
with oval shape.\cite{Adler:2001nb,Adler:2003kt} \ Plotted as 
a function of the transverse kinetic energy and scaled by the 
number of valence quarks of the hadron ($n=2$ for mesons and $n=3$ 
for baryons), the flow parameter $v_2$ exhibits a universal 
dependence.\cite{Adams:2003am} \ This has been interpreted
as evidence for a partonic origin of the observed flow pattern,
suggesting that the transverse expansion of the matter is 
generated during a phase in which it contains independent
quasi-particles with the quantum numbers of 
quarks\cite{Voloshin:2002wa,FMNB:03,Fries:2003kq,Greco:2003xt,MoVo:03} . 

The observed magnitude of the elliptic flow requires an early 
onset of the period during which the expansion is governed
by fluid dynamics (earlier than 1 fm/c after first impact) 
and nearly ideal fluid properties with a viscosity-to-entropy
density ratio $\eta/s \ll 
1$.\cite{Kolb:2003dz,Huovinen:2003fa,Hirano:2002hv,Nonaka:2006yn} \
This result, interpreted in a
quasi-particle picture, in turn requires that the mean free 
path in the medium must be extremely short, less than
the average particle spacing. A short mean free path
is also deduced from the strong suppression of the emission 
of hadrons with a transverse momentum $p_T$ of several GeV/c 
or more.\cite{Adcox:2002pe,Adler:2002xw} \ This phenomenon, 
commonly referred to as ``jet quenching'', is attributed to a 
large rate of energy loss of the energetic parton in the medium 
before it fragments into the observed 
hadron.\cite{Bjorken:1982tu,Thoma:1990fm,Wang:1991xy,Gyulassy:1993hr,Baier:1996kr,Zakharov:1997uu,Gyulassy:2000fs,Wiedemann:2000za,Baier:2001yt} \
Since the energy loss is inversely proportional to the mean free 
path of the parton, this observation again requires a very short 
mean free path or, equivalently, a large scattering cross section. 

There are other, purely theoretical reasons for believing 
that strongly interacting matter makes a rapid transition 
in a narrow temperature range around $T_c \approx 170$ MeV
from a hadronic resonance gas to a plasma whose thermodynamic
properties can be well described by quasiparticle excitations
with the quantum numbers of quarks and gluons. Phenomenological
attempts to describe the thermodynamic variables in terms of 
noninteracting quasiparticles with an effective thermal mass 
work astonishingly well for all temperatures down to 
$T_c$.\cite{Peshier:1999ww,Schneider:2001nf,Thaler:2003uz,Bluhm:2004xn} \
Rigorous resummation techniques using effective quasiparticle 
propagators for quarks and gluons are also quite successful
above $(1.5-2)T_c$.\cite{Blaizot:2003iq,Andersen:2003zk}

Even more compelling is the comparison of the
temperature dependence of the diagonal and off-diagonal quark 
flavor susceptibilities calculated on the lattice 
\cite{Gavai:2005sd} with expectations from a quasiparticle
picture. An especially sensitive quantity is the ratio between 
the baryon number-strangeness correlation $\langle{\cal B}S \rangle$
and the strangeness fluctuation $\langle S^2 \rangle$,\cite{Koch:2005vg} \ 
which rapidly changes from the behavior characteristic of a hadron gas 
to the temperature independent value of a quasiparticulate quark-gluon 
plasma at $T_c$.\cite{Majumder:2006nq} \ This result strongly suggests 
that the quasiparticles that carry flavor must have the same quantum 
numbers as quarks almost immediately above $T_c$ (for an alternate view, 
see Ref.~\citen{Liao:2005pa}). 

However, given the experimental evidence for the low viscosity and 
strong color opacity of the matter, the interpretation of the data 
in terms of the quasiparticle picture is problematic: In a 
quasiparticle picture the conjectured lower bound $\eta/s \geq 1/4\pi$ 
corresponds to an extremely short mean free path. Using the relation 
$\eta \approx n \lambda_f \bar{p}/2$ from standard 
kinetic theory, where $n$ denotes the particle density and 
$\lambda_f$ is the mean free path, as well as the equations 
$\bar{p} = 3T$ and $s\approx 4n$ holding for massless
particle, one finds $\lambda_f \geq (3\pi T/2)^{-1}$. This implies 
that at the lower viscosity bound the mean free path must not be 
larger than about half the average distance between quasiparticles.
This observation is corroborated by simulations of the parton
transport within the framework of the Boltzmann equation with
binary elastic scattering, which require cross sections up to 
twenty times larger than expected on the basis of perturbative 
QCD in order to reproduce the elliptic flow data.\cite{Molnar:2004yh} \
Therefore a collisional origin 
of such a low viscosity is not easily compatible with the notion 
that the medium is composed of well-defined quasiparticles.
The quasiparticle picture is further cast into doubt
by recent calculations of the spectral densities of correlators
of the stress-energy tensor in an exactly solvable, strongly 
coupled quantum field theory ($N=4$ supersymmetric Yang-Mills
theory).\cite{Teaney:2006nc,Kovtun:2006pf} \ These do not reveal
peak-like structures that can be attributed to quasiparticle 
excitations, with the sole exception of the hydrodynamical sound
mode. 

One possible resolution of the puzzle is to argue \cite{Gyulassy:2004zy}  
that the quark-gluon plasma near $T_c$ is actually strongly coupled.
This argument is bolstered by the fact
that results from solvable strongly coupled gauge theories are 
in qualitative agreement with certain aspects of QCD near $T_c$,
such as reduction of the energy density with respect to the 
ideal gas limit \cite{Gubser:1998nz}, and with experimentally 
obtained values of transport coefficients, such as the 
viscosity,\cite{Policastro:2001yc,Kovtun:2004de,Buchel:2004di} \ 
energy loss parameter,\cite{Liu:2006ug} \ and quark diffusion 
constant.\cite{Casalderrey:2006rq,Herzog:2006gh,Herzog:2006se} \ 
It has been argued \cite{Shuryak:2004tx,Gelman:2006xw,Liao:2005pa} 
that the microscopic structure of such a system is dominated by 
complex bound states of the elementary constituents. Note, however,
that the diagonal and off-diagonal quark flavor susceptibilities 
calculated on the lattice \cite{Gavai:2005sd} strongly constrain 
-- and in many cases rule out -- the existence of bound states of 
the elementary constituents above $T_C$. In addition, a 
recent calculation of the shear viscosity over entropy ratio 
in weakly coupled ${\cal N}=4$ supersymmetric Yang-Mills theory yielded
a result many times smaller than the corresponding weak-coupling
result in QCD.\cite{Huot:2006ys} \ This finding therefore may 
actually suggest that the ratio $\eta/s$ of QCD near the transition 
point is several times larger than the lower viscosity bound of 
$(4\pi)^{-1}$ for this quantity.\cite{Kovtun:2004de}

The other possible resolution is that the transport properties
of the quark-gluon plasma under the experimentally relevant 
conditions are not governed by collisional processes involving
perturbative interactions among elementary excitations, but by 
collective phenomena. This situation is not uncommon in plasmas,
where coherent fields can be spontaneously generated due to 
instabilities in the field equations in the presence
of the medium. The occurrence of unstable field modes is a 
familiar phenomenon in electromagnetic plasmas with a charged
particle distribution that is locally not fully equilibrated.
The most relevant of these for our purposes is the instability
discovered by Weibel, which arises when the momentum distribution
of charged particles is anisotropic.\cite{Weibel:1959}

It has been known for some time 
\cite{Mrowczynski:1988dz,Mrowczynski:1993qm,Mrowczynski:1996vh} 
that similar instabilities exist in quark-gluon plasmas with a 
parton momentum distribution that is not in thermal equilibrium.
As a result of these instabilities long-range color fields can be 
excited with amplitudes far above the thermal level. The generic 
nature of such color instabilities has been recognized only in recent 
years.\cite{Romatschke:2003ms,Arnold:2003rq} \ Most work exploring the 
consequences of these instabilities 
\cite{Randrup:2003cw,Arnold:2004ti,Mueller:2005un,Arnold:2004ih} has been 
focused on the early stage of the collision, when the momentum distribution 
is highly anisotropic and far from equilibrium. The fields generated 
by the instabilities drive the parton distribution rapidly toward 
local isotropy and thus toward the hydrodynamical 
regime.\cite{Mrowczynski:2005ki} \ However, the expansion
of the quark-gluon plasma under its own pressure ensures that the 
matter never reaches complete equilibrium, and thus the presence of
the color instabilities persists even during the period when the 
matter evolves by viscous hydrodynamical expansion, although the 
effect of collisions may weaken the color instabilities.\cite{Schenke:2006xu} \
Since the size of the deviation from kinetic equilibrium is proportional to 
the viscosity itself, color instabilities are especially important when the 
quark-gluon plasma is weakly coupled and the collisional shear
viscosity is large. By producing an anomalous contribution to the 
shear viscosity, extended color fields of a large amplitude present 
a mechanism that may be responsible for the observed small shear 
viscosity of the rapidly expanding quark-gluon plasma 
studied at RHIC.\cite{Asakawa:2006tc}

\section{Anomalous Viscosity: Heuristic Derivation}

Anomalous contributions to transport coefficients caused by the action
of turbulent fields are well known in electromagnetic plasmas. As we 
already mentioned, the condition for the spontaneous formation of 
extended electromagnetic fields in a plasma is the existence of 
instabilities in the field equations due to the interaction with 
the charged particles. This condition is satisfied in electromagnetic 
plasmas with an anisotropic momentum distribution.\cite{Weibel:1959} \ 
The treatment of the consequences for transport processes is based on 
the formalism of particle propagation in turbulent plasmas originally
developed by Dupree.\cite{Dupree:1966,Dupree:1968} \ The term {\em plasma
turbulence} refers to the spectral distribution of the field excitations, 
which follows a power law, in analogy to the Kolmogorov spectrum of vortex
excitations in a fluid with fully developed turbulence. Such plasmas 
are characterized by strongly excited random field modes in certain 
regimes of instability, which coherently deflect the charged particles 
and thus reduce the effective mean free path. The scattering by 
turbulent fields in electromagnetic plasmas is known to greatly reduce  
the heat conductivity \cite{Malone:1975,Okada:1978} and shear viscosity 
\cite{Abe:1980a,Abe:1980b} of the plasma and to increase the energy 
loss of charged particles propagating through it.\cite{Okada:1980} 

Following Abe and Niu \cite{Abe:1980b} the contribution from turbulent 
fields to plasma transport coefficients is called {\em anomalous}.
We shall see below that this designation is justified by the fact 
that the viscous corrections to the hydrodynamic energy-stress tensor
due to the turbulent fields are nonlinear in the velocity gradient, but
exhibit a sublinear dependence. As a consequence, the viscous effects 
in a quark-gluon plasma with a large imprinted velocity gradient are 
much smaller than expected from the usual linear theory. As the 
formation mechanism of quasi-thermal QCD matter in relativistic heavy
ion collisions naturally leads to a large longitudinal velocity 
gradient in the direction of the beam axis, the relevance of such
anomalous contributions to various transport coefficients, including
the shear viscosity, is not unexpected, especially during the early 
stages of the expansion.

Before describing the derivation of the anomalous shear viscosity 
$\eta_{\rm A}$ of an expanding quark-gluon plasma in detail, it is 
useful to give a heuristic argument for its dependence on the 
amplitude of turbulent plasma fields. This argument will also elucidate
the reason for the dominance of the anomalous viscosity at weak coupling. 
According to classical transport theory, the shear viscosity is given 
by an expression of the form \cite{Danielewicz:1984ww}
\begin{equation}
\eta \approx \frac{1}{3} n \bar{p}\lambda_{\rm f} , 
\label{eq:eta-class}
\end{equation}
where $n$ denotes the particle density, $\bar{p}$ is the thermal 
momentum, and $\lambda_{\rm f}$ the mean free path. For a weakly coupled 
quark-gluon plasma, $n \approx 5T^3$ and $\bar{p}\approx 3T$. The mean free 
path depends on the mechanism under consideration. The collisional shear
viscosity $\eta_{\rm C}$ is obtained by expressing the mean free path 
in terms of the transport cross section
\begin{equation}
\lambda_{\rm f}^{\rm (C)} = (n\,\sigma_{\rm tr})^{-1} .
\label{eq:lam-C}
\end{equation}
Using the perturbative QCD expression \cite{Danielewicz:1984ww}
\begin{equation}
\sigma_{\rm tr}\approx \frac{5g^4}{4\bar{p}^2}\ln\frac{\sqrt{4\pi}}{g}
\label{eq:sigma-tr}
\end{equation} 
for the transport cross section in a quark-gluon plasma yields the result 
\begin{equation}
\eta_C \approx \frac{T}{\sigma_{\rm tr}} 
\approx \frac{18\pi s}{25g^4\ln(\sqrt{4\pi}/g)} ,
\label{eq:eta-C-app}
\end{equation}
where we used the relation $s \approx 4n$ valid for ulrarelativistic
particles. This agrees parametrically with the result (\ref{eq:eta-C}) 
for the collisional shear viscosity in leading logarithmic 
approximation.\cite{Arnold:2000dr}\footnote{The logarithm in 
(\ref{eq:sigma-tr}) is the weak coupling limit of the function 
\cite{Danielewicz:1984ww} $I(\alpha_s)=(2\alpha_s+1)\ln(1+\alpha_s^{-1})-2$, 
which never becomes negative. In fact, $\alpha_s^2I(\alpha_s) \to 1/6$ 
for $\alpha_s\gg 1$, which suggests that $\eta/s$ approaches a constant 
in the strong coupling limit. Of course, the derivation of 
(\ref{eq:sigma-tr}) becomes invalid in this limit.}

The anomalous viscosity is determined by the same relation 
(\ref{eq:eta-class}) for $\eta$, but the mean free path is now 
obtained by counting the number of color field domains a thermal 
parton has to traverse in order to ``forget'' its original direction 
of motion. If we denote the field strength generically by ${\cal B}^a$ 
($a$ denotes the color index), a single coherent domain of size 
$r_{\rm m}$ causes a momentum deflection of the order of 
$\Delta p \sim gQ^a{\cal B}^ar_{\rm m}$, where $Q^a$ is the color charge
of the parton. If different field domains are uncorrelated, the mean 
free path due to the action of the turbulent fields is given by
\begin{equation}
\lambda_{\rm f}^{\rm (A)} = r_{\rm m}\langle(\bar{p}/\Delta p)^2\rangle
\sim \frac{\bar{p}^2}{g^2Q^2\langle{\cal B}^2\rangle r_{\rm m}} .
\label{eq:lam-A}
\end{equation}
The anomalous shear viscosity thus takes the form:
\begin{equation}
\eta_{\rm A} 
\sim \frac{n\,\bar{p}^3}{3g^2Q^2\langle{\cal B}^2\rangle r_{\rm m}}
\sim \frac{9sT^3}{4g^2Q^2\langle{\cal B}^2\rangle r_{\rm m}},
\label{eq:eta-A-app}
\end{equation}
which agrees with expression (16) of Ref.~\citen{Asakawa:2006tc}, 
if we identify $r_{\rm m}$ with the memory time $\tau_{\rm m}$ 
for relativistic partons. 

The argument now comes down to an estimate of the average field
intensity $\langle{\cal B}^2 \rangle$ and size $r_{\rm m}$ of a domain. 
We first note that the size is given by the characteristic wave 
length of the unstable field modes. Near thermal equilibrium, the
parameter describing the influence of hard thermal partons on the
soft color field modes is the color-electric screening mass 
$m_{\rm D}\sim gT$. Introducing a dimensionless parameter $\xi$ for 
the magnitude of the momentum space anisotropy \cite{Romatschke:2003ms},
the wave vector domain of unstable modes is $k^2 \leq \xi m_{\rm D}^2$
(see Appendix B). Thus $r_{\rm m}\sim \xi^{-1/2}(gT)^{-1}$. The 
exponential growth of the unstable soft field modes is saturated,
when the nonlinearities in the Yang-Mills equation become of the 
same order as the gradient term: $g|A|\sim k$, which implies 
that the field energy in the unstable mode is of the order of
$g^2\langle{\cal B}^2\rangle \sim k^4 \sim \xi^2 m_{\rm D}^4$. The 
denominator in (\ref{eq:eta-A-app}) thus has the characteristic
size, at saturation:
\begin{equation}
g^2Q^2\langle{\cal B}^2\rangle r_{\rm m} \sim \xi^{3/2} m_{\rm D}^3
\sim \xi^{3/2}(gT)^3 .
\label{eq:gB2r}
\end{equation}
Inserting this result into the expression (\ref{eq:eta-A-app}) gives
the following relation for the anomalous viscosity:
\begin{equation}
\eta_{\rm A} \sim \frac{s}{g^3\xi^{3/2}}.
\label{eq:eta-A-app2}
\end{equation}
We conclude that the anomalous viscosity will be smaller than the 
collisional viscosity, if the coupling constant $g$ is sufficiently 
small and the anisotropy parameter $\xi$ is sufficiently large.

\section{Diffusive Vlasov-Boltzmann Equation}

In order to present a more rigorous derivation of the anomalous viscosity
of a turbulent quark-gluon plasma, we first need to derive the appropriate
transport equation. The propagation of quasi-thermal 
partons in the presence of soft, locally coherent color fields and hard 
collisions among the partons is described by a Vlasov-Boltzmann equation 
\cite{Heinz:1983nx}:
\begin{equation}
v^\mu\frac{\partial}{\partial x^\mu} f({\mathbf r},{\mathbf p},t) 
+ g {\mathbf F}^a\cdot\nabla_p f^a({\mathbf r},{\mathbf p},t) 
+ C[f] = 0 \, .
\label{eq:Vlasov}
\end{equation}
Here $f({\mathbf r},{\mathbf p},t)$ denotes the usual parton distribution in 
phase space, which sums over all parton colors. $f^a({\mathbf r},{\mathbf p},t)$ 
denotes the color octet distribution function, which weights each parton 
with its color charge $Q^a$. Both, $f$ and $f^a$ can be defined in the 
semiclassical formalism \cite{Heinz:1984yq} as the moments of the distribution 
function ${\tilde f}({\mathbf r},{\mathbf p},Q,t)$ in an extended phase space 
that includes the color sector:
\begin{subequations}
\label{eq:fa}
\begin{eqnarray}
f({\mathbf r},{\mathbf p},t) 
&=& \int dQ\, {\tilde f}({\mathbf r},{\mathbf p},Q,t) \, ,  
\\
f^a({\mathbf r},{\mathbf p},t) 
&=& \int dQ\, Q^a {\tilde f}({\mathbf r},{\mathbf p},Q,t) \, .
\end{eqnarray}
\end{subequations}
In (\ref{eq:Vlasov}) we have used the covariant notation $v^\mu = p^\mu/p^0$ 
with $p^\mu=(E_p,{\mathbf p})$. ${\mathbf v}={\mathbf p}/E_p$ denotes the 
velocity of a parton with momentum ${\mathbf p}$ and energy $E_p$.  
Furthermore,
\begin{equation}
{\mathbf F}^a = {\cal E}^a + {\mathbf v}\times{\cal B}^a
\label{eq:Lorentz}
\end{equation}
denotes the color Lorentz force, and $C[f]$ stands for the collision term. 
(We will specify $C[f]$ later.) 

The color octet distribution $f^a$ satisfies a transport equation of its 
own, which couples it to phase space distributions of even higher color-SU(3) 
representations. In the vicinity of the equilibrium distribution, 
however, it makes sense to truncate the hierarchy at the level of
the color singlet and octet distributions and to retain only the 
lowest terms in the gradient expansion. We also note that the color 
octet distribution function vanishes in equilibrium, $f^a_0 = 0$, 
implying that $f^a$ is at least of first order in the perturbation.

 The transport equation for 
$f^a$ then reads:\cite{Heinz:1983nx,Heinz:1984yq}
\begin{equation}
v^\mu \frac{\partial f^a}{\partial x^\mu} + g f_{abc} A^b_\mu v^\mu f^c
+ \frac{g C_2}{N_c^2-1} {\mathbf F}^a\cdot\nabla_p f 
+ C^a[f,f^a] = 0,
\label{eq:Vlasov-8}
\end{equation}
where $C_2$ denotes the quadratic Casimir invariant of the color
representation of the thermal partons.

Before we linearize the Vlasov-Boltzmann equation (\ref{eq:Vlasov}),
we must rewrite it in a form applicable to the case of a turbulent 
quark-gluon plasma. In order to do so, we need to make additional
assumptions about the field distribution in the Vlasov force term.
Based on the arguments presented in Section II, we shall assume that the 
color field is turbulent, i.~e.\ random with a certain spatial and 
temporal correlation structure for fields at different space-time 
points. This assumption will allow us to rewrite the force term 
involving the color octet distribution function $f^a$ into a dissipative 
(Langevin) term acting on the color singlet distribution $f$.

We shall derive the diffusion term in two different ways. In this
Section, we will use linear response theory to calculate $f^a$ from
the color singlet distribution for given color fields. The diffusion term
is then obtained after substituting the result into the force term 
in (\ref{eq:Vlasov}) and choosing appropriate correlation functions
for the color fields. In Appendix A, we shall present another 
derivation based on the extended distribution function 
${\tilde f}({\mathbf r},{\mathbf p},Q,t)$. 
There we also discuss the difference between our approach and the 
one due to Dupree \cite{Dupree:1966} for turbulent abelian plasmas.
 
In order to resolve eq.~(\ref{eq:Vlasov-8}) for the color octet 
distribution, we Fourier transform the dependence on the space-time 
coordinate $x^\mu = (t,{\mathbf r})$:
\begin{equation}
f^a({\mathbf p},x) = \int \frac{d^4k}{(2\pi)^4} e^{-ik\cdot x} f^a({\mathbf p},k) .
\label{eq:Fourier}
\end{equation}
We allow for an arbitrary particle distribution in momentum space,
but neglect any space-time dependence of the singlet distribution 
$f({\mathbf p})$. Ignoring the collision term and, for the moment, the 
gauge connection associated with the space-time derivative, the solution 
of eq.~(\ref{eq:Vlasov-8}) is given by:\cite{Heinz:1985qe}
\begin{equation}
f^a({\mathbf p},k) = -ig \frac{C_2}{N_c^2-1} 
  (v\cdot k +i\epsilon)^{-1}
  {\mathbf F}^a(k)\cdot\nabla_p\, f({\mathbf p}) \, ,
\label{eq:fa-resk}
\end{equation}
where $v\cdot k \equiv v^\mu k_\mu = k^0-{\mathbf v}\cdot{\mathbf k}$.
The gauge connection has the effect of adding a path-ordered factor
\begin{equation}
U_{ac}(x,x') = P\, \exp\left( -\int_{x'}^x f_{abc}A^b_\mu\, dx^\mu \right) \, ,
\label{eq:U}
\end{equation}
which parallel transports the gauge fields from $x'$ to $x$. Returning  
to coordinate space, (\ref{eq:fa-resk}) then takes the form:
\begin{equation}
f^a({\mathbf p},x) = -ig \frac{C_2}{N_c^2-1} 
  \int \frac{d^4k}{(2\pi)^4} \int d^4x'\, U_{ab}(x,x')
  \frac{e^{ik\cdot(x'-x)}}{v\cdot k +i\epsilon}
  {\mathbf F}^b(x')\cdot\nabla_p\, f({\mathbf p}) .
\label{eq:fa-resx}
\end{equation}
Inserting this solution for $f^a$ into eq.~(\ref{eq:Vlasov}) for the singlet 
distribution function $f$, the Vlasov force term takes the following form:
\begin{eqnarray}
g {\mathbf F}^a(x)\cdot\nabla_p f^a({\mathbf p},x) 
&=& - \frac{ig^2C_2}{N_c^2-1} {\mathbf F}^a(x)\cdot\nabla_p
      \int \frac{d^4k}{(2\pi)^4} \int d^4x'\, U_{ab}(x,x')
\nonumber \\
&& \qquad\times  
  \frac{e^{ik\cdot(x'-x)}}{v\cdot k +i\epsilon}
  {\mathbf F}^b(x')\cdot\nabla_p\, f({\mathbf p}) .
\label{eq:VT-1}
\end{eqnarray}

We now need to invoke the argument that the soft color fields are 
turbulent and that their action on the quasi-thermal partons in 
(\ref{eq:VT-1}) can be described by taking an ensemble average, which
can be factorized in the form 
\begin{equation}
\langle F^a_i(x) U_{ab}(x,x') F^b_j(x') f({\mathbf p}) \rangle 
= \langle F^a_i(x) U_{ab}(x,x') F^b_j(x') \rangle \bar{f}({\mathbf p}) ,
\label{eq:ens-avg}
\end{equation}
where $\bar{f} \equiv \langle f \rangle$. We furthermore assume that 
the correlation functions of fields at different space-time points 
$x$ and $x'$ depend only on $|x-x'|$ and fall off rapidly with 
correlation time $\tau$ and correlation length $\sigma$:
\begin{subequations}
\label{eq:F-correl}
\begin{eqnarray}
\langle {\cal E}^a_i(x) U_{ab}(x,x') {\cal E}^b_j(x') \rangle 
&=& \langle {\cal E}^a_i {\cal E}^a_j \rangle\, 
  \Phi^{\rm (el)}_\tau\left(|t-t'|\right)\, 
  \tilde\Phi^{\rm (el)}_\sigma\left(|{\mathbf x}-{\mathbf x}'|\right)\, ,
\\
\langle {\cal B}^a_i(x) U_{ab}(x,x') {\cal B}^b_j(x') \rangle 
&=& \langle {\cal B}^a_i {\cal B}^a_j \rangle\, 
  \Phi^{\rm (mag)}_\tau\left(|t-t'|\right)\, 
  \tilde\Phi^{\rm (mag)}_\sigma\left(|{\mathbf x}-{\mathbf x}'|\right)\, .
\end{eqnarray}
\end{subequations}
Examples satisfying these assumptions are the Gaussian correlators
\begin{subequations}
\label{eq:Phi-corr}
\begin{eqnarray}
\Phi^{\rm (el/mag)}_\tau\left(|t-t'|\right) 
&=&  \exp\left[(t-t')^2/2\tau_{\rm el/mag}^2\right]\, , 
\\
\tilde\Phi^{\rm (el/mag)}_\sigma\left(|{\mathbf x}-{\mathbf x}'|\right)
&=&  \exp\left[({\mathbf x}-{\mathbf x}')^2/2\sigma_{\rm el/mag}^2\right] \, .
\end{eqnarray}
\end{subequations}
Finally, we assume that the color-electric and -magnetic fields are 
uncorrelated: 
$\langle{\cal E}^a_i(x) U_{ab}(x,x') {\cal B}^b_j(x') \rangle = 0$. 

The reality of the correlation functions (\ref{eq:Phi-corr}) and their 
symmetry with respect to exchange of the two space-time arguments express 
the chaotic nature of the plasma. It is here where the reversibility of
the mean field dynamics is explicitly broken and the dissipative nature
of the turbulent plasma is introduced. The existence of plasma instabilities
and the associated exponential growth of the unstable modes, which 
correspond to the presence of positive Lyapunov exponents in the coupled
field-particle system, forms the essential physical basis for our argument.

Since the right-hand side of (\ref{eq:VT-1}) must be real, under 
the conditions outlined above only the imaginary part of the propagator
\begin{equation}
{\rm Im}\, \frac{1}{v\cdot k +i\epsilon} = - \pi \delta(v\cdot k)
\label{eq:Im-part}
\end{equation}
can contribute. Performing first the integral over $k^0$, then the integral
over ${\mathbf k}$ and finally the integral over ${\mathbf x}'$, we obtain:
\begin{eqnarray}
\langle g{\mathbf F}^a\cdot\nabla_p\, f^a \rangle
&=& - \frac{g^2C_2}{N_c^2-1} \left[ \tau_{\rm m}^{\rm el}
      \langle{\cal E}^a_i {\cal E}^a_j \rangle\, 
      \frac{\partial^2}{\partial p_i \partial p_j} 
    + \tau_{\rm m}^{\rm mag} \langle{\cal B}^a_i {\cal B}^a_j \rangle \,
      ({\mathbf v}\times\nabla_p)_i ({\mathbf v}\times\nabla_p)_j \right] 
    \bar{f}({\mathbf p})
\nonumber \\ 
&\equiv & - \nabla_p \cdot D({\mathbf p}) \cdot \nabla_p\, 
   \bar{f}({\mathbf p}) \, .
\label{eq:VT-2}
\end{eqnarray}
Here we made use of the identity 
$({\mathbf v}\times{\cal B})\cdot\nabla_p 
= -{\cal B}\cdot({\mathbf v}\times\nabla_p)$ and introduced 
the memory time
\begin{equation}
\tau_{\rm m}^{\rm el/mag} 
= \frac{1}{2} \int_{-\infty}^{\infty} dt' \, 
  \Phi^{\rm (el/mag)}_\tau\left(|t-t'|\right) \,
  \tilde\Phi^{\rm (el/mag)}_\sigma
        \left(|{\mathbf v}(t-t')|\right)\, .
\label{eq:tau-mem}
\end{equation}

The precise matrix structure of the correlators (\ref{eq:F-correl}) may 
differ between the various stages of a relativistic heavy ion collision. 
Initially, color-magnetic field modes exhibit the strongest growth rate 
\cite{Rebhan:2004ur}, but in the later turbulent steady-state regime 
all spatial components of the color field modes populated by the 
instabilities are expected to be of comparable 
strength.\cite{Arnold:2005ef,Arnold:2005qs} \ 
For completeness, we shall discuss both scenarios. 

When we consider only the color-magnetic fields initially generated 
by the plasma instability, which point in a transverse direction, we 
can expect the ensemble average to be transverse in space:
\begin{equation}
{\rm (A)}\qquad
\langle{\cal B}^a_i {\cal B}^a_j \rangle 
= \frac{1}{2}(\delta_{ij}-\delta_{iz}\delta_{jz}) 
  \langle{\cal B}^2\rangle\, ;
\qquad
\langle{\cal E}^a_i {\cal E}^a_j \rangle \approx 0 \, .
\label{eq:B-corr-a}
\end{equation}
We shall call this case {\em Scenario A}.
In the late turbulent phase all spatial components of the field 
correlators are of approximately equal size, and we write:
\begin{equation}
{\rm (B)}\qquad
\langle{\cal B}^a_i {\cal B}^a_j \rangle 
= \frac{1}{3}\delta_{ij}\langle{\cal B}^2\rangle\, ;
\qquad
\langle{\cal E}^a_i {\cal E}^a_j \rangle 
= \frac{1}{3}\delta_{ij}\langle{\cal E}^2\rangle\, .
\label{eq:B-corr-b}
\end{equation}
We shall call this case {\em Scenario B}. Employing the notation 
$-i{\mathbf p}\times\nabla_p = {\mathbf L}^{(p)}$ for the generator of 
rotations in momentum space, we can write the diffusive term in the 
transverse color-magnetic field dominated Scenario A as
\begin{equation}
\nabla_p\cdot D \cdot\nabla_p 
= - \frac{g^2 C_2}{2(N_c^2-1)E_p^2}\langle{\cal B}^2 \rangle\, 
    \tau_{\rm m}^{\rm mag} 
    \left[({\mathbf L}^{(p)})^2 - (L^{(p)}_z)^2\right] \, ,
\label{eq:D-1a}
\end{equation}
and in the Scenario B with isotropically turbulent color-electric and 
color-magnetic field excitations as
\begin{equation}
\nabla_p\cdot D \cdot\nabla_p 
= \frac{g^2 C_2}{3(N_c^2-1)} 
  \left[ \langle{\cal E}^2\rangle\,\tau_{\rm m}^{\rm el}\,\nabla_p^2 
  - \langle{\cal B}^2 \rangle\,\tau_{\rm m}^{\rm mag}\,
    \frac{({\mathbf L}^{(p)})^2}{E_p^2} \, \right] .
\label{eq:D-1b}
\end{equation}
Note that the operator $\nabla_p^2$ associated with random color-electric 
fields yields a nonvanishing contribution when acting on the equilibrated 
momentum distribution $f_0({\mathbf p})$. This is different for the operator 
associated with random color-magnetic fields, $({\mathbf L}^{(p)})^2$, 
which only yields a nonvanishing contribution when acting on the 
anisotropic part of the momentum distribution given by $f_1({\mathbf p})$.
This is not surprising, because randomly distributed electric fields are 
well known to lead to an increase in the average energy of the plasma 
particles \cite{Fermi:1949ee}, corresponding to a heating of the plasma. 
In contrast, color-magnetic fields only contribute to the isotropization 
of the momentum distribution and do not cause plasma heating.

\section{Linear Response Theory}

We now have motivated the replacement of the Vlasov-Boltzmann equation
(\ref{eq:Vlasov}) by Dupree's ensemble averaged, diffusive Vlasov-Boltzmann 
equation
\begin{equation}
v^\mu\frac{\partial}{\partial x^\mu} \bar{f} 
- \nabla_p\cdot D \cdot\nabla_p \bar{f} + \langle C[f]\rangle = 0 \, .
\label{eq:DVBE}
\end{equation}
We next need to expand this equation up to linear terms in the ensemble
averaged perturbation $\delta\bar{f}$ of the parton distribution. The 
drift term gives rise to gradients of the collective variables $T$ and 
$u^\mu$, which are considered to be of the same size as terms linear in 
$\delta\bar{f}$; the collision term vanishes at equilibrium and
contributes only at first order in $\delta\bar{f}$; and the Vlasov term 
will require special consideration. 

The general linear response (Chapman-Enskog) formalism assumes 
a small perturbation of the thermal equilibrium distribution
\begin{equation}
f_0({\mathbf p}) = \left(e^{\beta u\cdot p}\mp 1\right)^{-1} ,
\label{eq:f0}
\end{equation}
where $\beta=1/T$ denotes the inverse temperature and the upper/lower
sign applies to bosons/fermions, respectively.
Here $u^{\mu}$ denotes the four-velocity of the equilibrated medium 
and $u\cdot p \equiv u^\mu p_\mu$. Using the relation 
$T\,\partial f_0/\partial(u\cdot p)=-f_0(1\pm f_0)$, one expresses 
the perturbed distribution function in the form:
\begin{equation}
f({\mathbf p},{\mathbf r}) 
= f_0({\mathbf p}) + \delta f({\mathbf p},{\mathbf r})
= f_0({\mathbf p}) [1+f_1({\mathbf p},{\mathbf r})(1\pm f_0({\mathbf p}))] ,
\label{eq:f0-f1}
\end{equation}
where the function $f_1({\mathbf p},{\mathbf r})$ can be considered as (minus) 
the change in the argument of the equilibrium distribution:\cite{Landau10}
\begin{equation}
f({\mathbf p},{\mathbf r}) \approx 
\left(e^{\beta u\cdot p-f_1({\mathbf p},{\mathbf r})}\mp 1\right)^{-1} .
\label{eq:f1arg}
\end{equation}

While the relativistic dissipative hydrodynamics can be cast in a 
manifestly covariant formalism \cite{Israel:1976tn,Israel:1979wp}, 
this is not necessary for the derivation of the transport coefficients,
which are Lorentz invariants. It is thus convenient \cite{Chen:2006ig} 
to work in the local rest frame of the fluid, i.~e.\ in the frame where 
${\mathbf u}(x)=0, u^0=1$ and thus $u\cdot p = p^0 = E_p$ at the 
considered space-time point $x$. This choice also implies the 
relation $\partial u^0/\partial x^\mu = 0$.

In the derivation of transport coefficients one assumes that the 
particle distribution is slowly varying, so that the deviation from
the equilibrium distribution is homogeneous in space and proportional
to gradients of the equilibrium parameters. For the shear viscosity one 
uses the local perturbation\footnote{Our definition makes 
$\bar\Delta({\mathbf p})$ dimensionless. To compare with 
Ref.~\citen{Arnold:2000dr}, identify $\bar\Delta(p)=-\chi(p)E_p/{\mathbf p}^2$. 
In order to compare with Ref.~\citen{Landau10}, identify 
$f_1(p)=\chi(p)/T$ and $\bar\Delta(p)=-g(p)T/E_p$. Also note that 
$(\nabla u)_{ij}$ projects out the traceless part of $p_i p_j$.} 
\begin{equation}
f_1({\mathbf p},{\mathbf r}) 
= - \frac{\bar\Delta(p)}{E_p T^2} p_i p_j (\nabla u)_{ij} ,
\label{eq:f1}
\end{equation}
where $\bar\Delta(p)$ is a scalar function of the momentum $p$, which 
measures the magnitude of the deviation from equilibrium, 
and $(\nabla u)_{ij}$ denotes the traceless 
symmetrized velocity gradient:
\begin{equation}
(\nabla u)_{ij} \equiv \frac{1}{2}(\nabla_iu_j+\nabla_ju_i)
  -\frac{1}{3}\delta_{ij}\nabla\cdot{\mathbf u} ,
\label{eq:nabla-u-ij}
\end{equation}
For Bjorken's \cite{Bjorken:1982qr} boost invariant flow field 
$u_z = z/\tau$ we have
\begin{equation}
(\nabla u)_{ij} = \frac{1}{3\tau} {\rm diag}(-1,-1,2) .
\label{eq:nabla-u}
\end{equation}

The connection to the shear viscosity is made by comparing the 
microscopic definition of the stress tensor 
\begin{equation}
T_{ik} = \int \frac{d^3p}{(2\pi)^3E_p}\, p_ip_k f({\mathbf p},{\mathbf r}) 
\label{eq:T-tensor}
\end{equation}
with the macroscopic definition of the viscous stress:
\begin{equation}
T_{ik} = T_{ik}^{(0)} + \delta T_{ik} 
= P\delta_{ik} + \varepsilon u_iu_k 
  - 2\eta(\nabla u)_{ik} - \zeta\delta_{ik}\nabla\cdot{\mathbf u} ,
\label{eq:T-eta}
\end{equation}
where $\varepsilon$ and $P$ are the equilibrium energy density
and pressure, and where $\eta$ and $\zeta$ denote the shear and bulk 
viscosity, respectively.

Combining equations (\ref{eq:f0-f1}), (\ref{eq:f1}) and (\ref{eq:T-tensor}) 
one finds
\begin{eqnarray}
\delta T_{ik} &=& \int \frac{d^3p}{(2\pi)^3E_p}\, 
  p_ip_k f_1({\mathbf p}) f_0({\mathbf p})(1\pm f_0({\mathbf p}))
\nonumber \\
&=& \frac{1}{T} (\nabla u)_{mn} 
    \int \frac{d^3p}{(2\pi)^3E_p^2}\, p_ip_k \bar\Delta(p)
         p_m p_n \frac{\partial f_0}{\partial E_p} 
\nonumber \\
&=& - \eta (\nabla u)_{mn} [\delta_{im}\delta_{kn} 
      + \delta_{in}\delta_{km} + \delta_{ik}\delta_{mn}] ,
\label{eq:dTik-1}
\end{eqnarray}
with the shear viscosity coefficient
\begin{equation}
\eta = - \frac{1}{15T} \int \frac{d^3p}{(2\pi)^3}\, 
   \frac{{\mathbf p}^4}{E_p^2}\bar\Delta(p)\frac{\partial f_0}{\partial E_p} .
\label{eq:eta-Delta}
\end{equation}

For the special case (\ref{eq:nabla-u}) of boost-invariant longitudinal 
flow, the perturbation of the equilibrium distribution takes the form
\begin{equation}
f_1({\mathbf p}) = - \frac{\bar\Delta(p)}{E_p T^2}\, |\nabla u|
                     \left(p_z^2-\frac{{\mathbf p}^2}{3}\right) ,
\label{eq:f1-Qzz}
\end{equation}
with (note the normalization):
\begin{equation}
|\nabla u| 
\equiv \left[\frac{3}{2}(\nabla u)_{ij}(\nabla u)_{ji}\right]^{1/2}
= \frac{1}{\tau} \, ,
\label{eq:grad-u}
\end{equation}
and the anisotropy of the stress tensor (\ref{eq:dTik-1}) 
is\cite{Teaney:2003pb,Heiselberg:1995sh}
\begin{equation}
2\delta T_{xx} = 2\delta T_{yy} = - \delta T_{zz}
= \frac{4}{3}\, \eta |\nabla u| \, .
\label{eq:Tzz}
\end{equation}

\section{Shear Viscosity}

\subsection{Drift Term}

We begin with the evaluation of the drift term. The dominant
contribution comes from the $r$-dependence of the local equilibrium 
distribution. Because the form of this term does not depend on ensemble
averaging, we shall omit the ``bar'' symbol in this section. Using a dot 
to denote the partial time derivative, one finds in the local rest frame 
(recall that ${\mathbf v}={\mathbf p}/E_p$):
\begin{equation}
v^\mu\frac{\partial}{\partial x^\mu} f_0({\mathbf p}) 
= - f_0(1\pm f_0) \left[ \dot\beta E_p - \beta\dot{{\mathbf u}}\cdot{\mathbf p}
           + {\mathbf v}\cdot\nabla\beta 
           - \beta{\mathbf v}\cdot\nabla({\mathbf u}\cdot{\mathbf p}) \right] .
\label{eq:drift-1}
\end{equation}
This expression can be simplified with the help of the energy-momentum 
conservation law in the presence of color fields:
\begin{equation}
\frac{\partial T^{\mu\nu}}{\partial x^\nu} = F^{a\mu\nu}j^a_\nu \, .
\label{eq:em-cons}
\end{equation}
With the help of the expression for $T^{\mu\nu}$ in terms 
of the momentum space distribution function, we find:
\begin{eqnarray}
\frac{\partial T^{\mu\nu}}{\partial x^\mu} 
&=& \int\frac{d^3p}{(2\pi)^3E_p}\, p^\mu p^\nu 
        \frac{\partial f_0}{\partial x^\mu}
\nonumber \\
&=& - \int\frac{d^3p}{(2\pi)^3E_p}\, p^\mu p^\nu
      f_0(1\pm f_0) \left( E_p \frac{\partial\beta}{\partial x^\mu}
     - \beta{\mathbf p}\cdot\frac{\partial{\mathbf u}}{\partial x^\mu} \right)
\nonumber \\
&=& \frac{\partial\beta}{\partial x^\mu} 
         \frac{\partial T^{\mu\nu}}{\partial\beta}
   - \frac{\partial{\mathbf u}}{\partial x^\mu}\cdot 
         \beta\frac{\partial}{\partial\beta} 
         \int\frac{d^3p}{(2\pi)^3E_p^2}\, p^\mu p^\nu {\mathbf p} f_0(p) .
\label{eq:partialT-1}
\end{eqnarray}
Using the expression (\ref{eq:fa-resx}) for the color octet distribution 
induced by the color field and following the same arguments that led to
the relation (\ref{eq:VT-2}), we obtain for the ensemble average of the
right-hand side of (\ref{eq:em-cons}):
\begin{eqnarray}
\langle F^{a\mu\nu}(x)j^a_\nu(x) \rangle
&=& g \int \frac{d^3p}{(2\pi)^3E_p} \langle F^{a\mu\nu}p_\nu 
    f^a({\mathbf p},x) \rangle 
\nonumber \\
&=& - \frac{g^2C_2}{N_c^2-1} \int \frac{d^3p}{(2\pi)^3E_p}
      p_\nu\tau_{\rm m}\langle F^{a\mu\nu}{\mathbf{F}}^a\rangle\cdot\nabla_p \,
      \bar{f}({\mathbf p})
\label{eq:Fj}
\end{eqnarray}
The time-like component ($\nu=0$) of the expressions (\ref{eq:partialT-1})
and (\ref{eq:Fj}) are easily evaluated to yield the relation
\begin{equation}
\dot{\beta}\frac{\partial\varepsilon}{\partial\beta}
  - (\nabla\cdot{\mathbf u})\beta\frac{\partial P}{\partial\beta} 
= \langle {\cal E}^a\cdot{\mathbf j}^a \rangle 
= \frac{m_{\rm D}^2}{3} \langle {\cal E}^2 \rangle \tau_{\rm m}^{\rm el} \, ,
\label{eq:partialT-2}
\end{equation}
where $m_{\rm D}$ is the Debye screening mass defined as\footnote{In the
notation of Ref.~\citen{Asakawa:2006tc} the coefficient of $g^2T^2$is 
given by the expression $\frac{N_c\nu'_2\zeta(2)}{(N_c^2-1)\pi^2}$.}
\begin{equation}
m_{\rm D}^2 
= - \frac{g^2C_2}{N_c^2-1} \int \frac{d^3p}{(2\pi)^3E_p}
    {\mathbf p}\cdot\nabla_p\, \bar{f}({\mathbf p})
= \left(\frac{N_c}{3}+\frac{N_f}{6}\right) g^2T^2 \, ,
\label{eq:mD}
\end{equation}
and the final expression holds for a noninteracting plasma of massless
quarks and gluons in thermal equilibrium. The space-like components yield,
after some calculations:
\begin{equation}
(\nabla\beta-\beta\dot{{\mathbf u}})\frac{\partial P}{\partial\beta} 
= - \langle {\cal E}^a j^{a0} + {\mathbf j}^a\times{\cal B}^a \rangle 
= -g \int \frac{d^3p}{(2\pi)^3} 
     \langle {\mathbf F}^a f^a({\mathbf p}) \rangle
= 0 \, .
\label{eq:partialT-3}
\end{equation}
The right-hand side vanishes by virtue of (\ref{eq:fa-resx}), because the
momentum integral reduces to a surface term at infinity.

We can now use the relations (\ref{eq:partialT-2}) and (\ref{eq:partialT-3}) 
to eliminate the time derivatives $\dot\beta$ and $\dot{\mathbf u}$ from 
the drift term (\ref{eq:drift-1}), obtaining:
\begin{eqnarray}
v^\mu\frac{\partial}{\partial x^\mu} f_0({\mathbf p}) 
&=& f_0(1\pm f_0) \left[ \frac{({\mathbf p}\cdot\nabla)({\mathbf u}\cdot 
    {\mathbf p})}{E_p T}
    - \frac{m_{\rm D}^2\langle{\cal E}^2\rangle\tau_{\rm m}^{\rm el}E_p}
           {3 T^2(\partial\varepsilon/\partial T)} \right.
\nonumber \\
& & \qquad\qquad\qquad \left.
    - \frac{\partial P/\partial T}{T(\partial\varepsilon/\partial T)}
      E_p(\nabla\cdot{\mathbf u}) \right] \, . 
\label{eq:drift-2}
\end{eqnarray}
The first term in the square brackets can be expressed in terms of the 
traceless velocity gradient (\ref{eq:nabla-u-ij}) and a term which can
be combined with the last term in the brackets, yielding
\begin{equation}
\left(\frac{{\mathbf p}^2}{3E_p^2} 
- \frac{\partial P/\partial T}{\partial\varepsilon/\partial T}\right) 
  \frac{E_p}{T} (\nabla\cdot{\mathbf u}) \, .
\label{eq:bulk-visc}
\end{equation}
For a gas of massless noninteracting partons, $\varepsilon=3P$ and 
${\mathbf p}^2=E_p^2$, causing the term proportional to the divergence of 
the collective velocity to vanish, in accordance with the expectation that
the bulk viscosity of a scale invariant system must be zero. Thus our final 
expression for the drift term is:
\begin{equation}
v^\mu\frac{\partial}{\partial x^\mu} f_0({\mathbf p}) 
= f_0(1\pm f_0) \left[ \frac{p_ip_j}{E_p T}(\nabla u)_{ij}
    - \frac{m_{\rm D}^2\langle{\cal E}^2\rangle\tau_{\rm m}^{\rm el} E_p}
           {3 T^2(\partial\varepsilon/\partial T)} \right] \, .    
\label{eq:drift-3}
\end{equation}

Equation (\ref{eq:partialT-2}) describes the heating of the plasma by 
the coherent color field. The coefficient of $\langle{\cal E}^2\rangle$,
\begin{equation}
\sigma_{\rm A} = \frac{1}{3} m_{\rm D}^2 \tau_{\rm m}^{\rm el} \, ,
\label{eq:c-cond}
\end{equation}
is the effective color conductivity of the turbulent plasma (see Appendix B 
of Ref.~\citen{Heinz:1985qe}). This expression for the conductivity differs 
from the collisional conductivity obtained for a plasma in equilibrium in the 
absence of coherent color fields.\cite{Selikhov:1993ns,Arnold:1998cy} \
In analogy to the anomalous shear viscosity $\eta_{\rm A}$ of the turbulent 
plasma, which is the main object of this article, $\sigma_{\rm A}$ may be
called the {\em anomalous} color conductivity.

\subsection{Force Term}

Next we come to the diffusive Vlasov force term in (\ref{eq:DVBE}).
Because they are associated with different operators in momentum space
(see eq.~(\ref{eq:D-1b})), the color-electric and color-magnetic 
contributions to the diffusion term need to be considered separately.
Since $f_0$ depends solely on $E_p=|{\mathbf p}|$ and hence 
${\mathbf L}^{(p)} f_0 = 0$, the color-magnetic contribution to the 
diffusion term affects only the deviation of the particle distribution 
from equilibrium. On the other hand, the color-electric contribution to 
the diffusion term affects the 
full momentum space distribution, because $\nabla_p^2\, f_0 \neq 0$.
The difference is easy to understand: color-magnetic fields only change
the direction of the momentum of a parton, leading to a rearrangement of
particles within the equilibrium distribution, but not to a modification 
of the distribution itself. Color-electric fields, on the other hand, 
accelerate partons and thus lead, on average, to a heating of the
thermal distribution. In the macroscopic formulation, this difference 
is related to the fact that color-electric fields induce a color current 
in the parton distribution, which interacts dissipatively with the 
color-electric field. We have already discussed this effect at the 
end of the previous section. 

We first consider the diffusion operator (\ref{eq:D-1a}) for Scenario A. 
Since the angular dependence of the perturbation $f_1({\mathbf p})$ in 
(\ref{eq:f1}) has the form of a quadrupole in momentum space, it is an 
eigenfunction of the magnetic diffusion operator. For the perturbation 
(\ref{eq:f1-Qzz}) associated with the Bjorken flow, $f_1({\mathbf p}) 
\sim Y_{20}(\hat{\mathbf p})$, implying that
\begin{equation}
\left[(L^{(p)})^2 - (L^{(p)}_z)^2\right] f_1({\mathbf p}) 
= \left[(L^{(p)})^2\right] f_1({\mathbf p}) 
= 6\, f_1({\mathbf p}) .
\label{eq:L2f}
\end{equation}
The color-magnetic part of the diffusion term in Scenario A thus
takes the form
\begin{equation}
\nabla_p\cdot D^{\rm mag} \cdot\nabla_p \bar{f}({\mathbf p})
= \frac{3\, C_2\,\bar\Delta\, g^2 \langle{\cal B}^2 \rangle\, 
        \tau_{\rm m}^{\rm mag}}{(N_c^2-1)E_p^3T^2}\,  
  f_0 (1\pm f_0) p_i p_j (\nabla u)_{ij} \, .
\label{eq:diffus-mag}
\end{equation}

The diffusion term for scenario B yields (with $p=|{\mathbf p}|$):
\begin{eqnarray}
\nabla_p\cdot D \cdot\nabla_p \bar{f}({\mathbf p})
&=& \frac{C_2\, g^2\langle{\cal E}^2 \rangle\, \tau_{\rm m}^{\rm el}}
         {3(N_c^2-1)p}\, \frac{\partial^2}{\partial p^2} 
         [p f({\mathbf p})] 
\nonumber \\
&& + \frac{2C_2\bar\Delta g^2}{(N_c^2-1)E_pT^2} 
     \left(\frac{\langle{\cal E}^2 \rangle\, \tau_{\rm m}^{\rm el}}{p^2}
           +\frac{\langle{\cal B}^2 \rangle\, \tau_{\rm m}^{\rm mag}}{E_p^2}
     \right) 
\nonumber \\ 
&& \qquad\qquad\qquad
   \times f_0 (1\pm f_0) p_ip_j (\nabla u)_{ij} \, .
\label{eq:diffus-el}
\end{eqnarray}
The rotationally symmetric part of the first term describes the heating 
of the parton distribution by the turbulent electric fields discussed 
in conjunction with the anomalous color conductivity (\ref{eq:c-cond}). 
The anisotropic part of the first term, together with the second term 
describes the angular diffusion of the parton distribution, which leads 
to an anomalous viscosity. These terms have the same structure as the 
force term in Scenario A except that color-electric as well as
color-magnetic fields contribute.

\subsection{Collision Term}

In the evaluation of the collision term 
\begin{eqnarray}
C[f]({\mathbf p}) &=& \frac{1}{4E_p} 
  \int \frac{d^3k d^3p' d^3k'}{(2\pi)^5 8 E_k E'_p E'_k}
  \delta^4(p+k-p'-k') \left|\sum_\alpha M_\alpha(p,k,p',k')\right|^2
\nonumber \\ 
&& \qquad \times
  \left( f({\mathbf p})f({\mathbf k})
         [1\pm f({\mathbf p}')][1\pm f({\mathbf k}')] \right.
\nonumber \\ & & \qquad \qquad \left.
       - f({\mathbf p}')f({\mathbf k}')
         [1\pm f({\mathbf p})][1\pm f({\mathbf k})] \right)\, 
\label{eq:coll}
\end{eqnarray}
we follow Arnold {\em et al.}\cite{Arnold:2000dr} \ 
Note that (\ref{eq:coll}) involves an implicit summation over parton 
flavors and helicities. Since the collision term vanishes at equilibrium 
owing to detailed balance, the leading contribution is linear in 
$f_1({\mathbf p})$. In first approximation, the collision term thus 
gives rise to a linear integral operator of the form
\begin{equation}
I[f_1]({\mathbf p}) 
=  \int \frac{d^3k}{(2\pi)^3} d\sigma_{12} v_{\rm rel} 
   f_0({\mathbf p})f_0({\mathbf k}) 
   \left[ f_1({\mathbf p}) + f_1({\mathbf k})
         - f_1({\mathbf p}') - f_1({\mathbf k}') \right] \, ,
\label{eq:coll-int}
\end{equation}
where
\begin{equation}
d\sigma_{12} 
= \frac{1}{8\sqrt{p\cdot k}} \frac{d^3p' d^3k'}{16\pi^2 E'_p E'_k}
  \delta^4(p+k-p'-k') \left|\sum_\alpha M_\alpha\right|^2
         [1\pm f_0({\mathbf p}')][1\pm f_0({\mathbf k}')] 
\label{eq:cross}
\end{equation}
denotes the differential cross section for the scattering process 
${\mathbf p},{\mathbf k}\to {\mathbf p}',{\mathbf k}'$ and $v_{\rm rel}$ 
is the relative velocity. We also note that ensemble average of the
collision term required in the diffusive Vlasov-Boltzmann equation 
(\ref{eq:DVBE}) simply translates into the averaged distribution function
$\bar{f}_1$ in the linearized collision term (\ref{eq:coll-int}):
$\langle I[f_1]\rangle = I[\bar{f}_1]$.

For the leading logarithmic limit of the collisional viscosity, it is
sufficient to use the contributions to the scattering matrix element
$M_\alpha$ with the highest infrared divergence. Using Mandelstam 
variables $s,t,u$, the squared matrix element for one-gluon exchange
processes (averaged over initial-state and summed over final-state
quantum numbers) is
\begin{equation}
\left|\bar{M}_{(ab)}\right|^2 = \frac{4 g^4}{N_c^2-1} \frac{s^2}{t^2}
  C_2^{(a)} C_2^{(b)} \, ,
\label{eq:M-gluon}
\end{equation}
where $a,b$ denote the quantum numbers of the scattering partons,
and $C_2^{(a,b)}$ are the quadratic Casimir operators for their color 
multiplet. The squared matrix element for the quark annihilation 
process is
\begin{equation}
\left|\bar{M}\right|^2 = \frac{4 g^4}{N_c} 
  \left( \frac{u}{t} + \frac{t}{u} \right) (C_2^{\rm (f)})^2 \, .
\label{eq:M-quark}
\end{equation}
Finally, the Compton scattering process doubles the contribution 
from quark annihilation.\cite{Arnold:2000dr} \

\subsection{Anomalous Viscosity}

We are now ready to calculate the coefficient of shear viscosity.
We begin by ignoring the collision term and calculate only the 
contribution of the diffusive Vlasov term, i.~e.\ the anomalous 
shear viscosity. Equating the first term in (\ref{eq:drift-3}) with 
the right-hand side of (\ref{eq:diffus-mag}) for Scenario A we obtain
\begin{equation}
\bar\Delta(p) = \frac{(N_c^2-1)E_p^2 T}
   {3C_2\,g^2\langle{\cal B}^2 \rangle\,\tau_{\rm m}^{\rm mag}} \, .
\label{eq:Delta}
\end{equation}
We note that the diffusive Vlasov equations for quarks and gluons 
decouple in the absence of collisions, causing the function 
$\bar\Delta(p)$ to take different values for quarks and 
gluons. Inserting (\ref{eq:Delta}) into the relation (\ref{eq:eta-Delta}) 
between $\eta$ and $\bar\Delta$ and assuming massless partons, the 
desired expression for the anomalous shear viscosity due to the action
of the coherent color-magnetic fields on massless partons is found to be:
\begin{equation}
\eta_{\rm A} 
= \frac{N_c^2-1}
         {15\pi^2C_2g^2\langle{\cal B}^2 \rangle\,\tau_{\rm m}^{\rm mag}}
    \int_0^\infty dp\,p^5 f_0(p) \, ,
\label{eq:eta-A1}
\end{equation}
where a sum over parton species and helicities is implied. After performing 
the momentum space integration, one obtains the following anomalous 
viscosities for gluons and quarks:
\begin{subequations}
\label{eq:eta-A2}
\begin{equation}
\eta_{\rm A}^{(g)} = \frac{16\zeta(6)(N_c^2-1)^2}{\pi^2N_c}
   \frac{T^6}{g^2\langle{\cal B}^2 \rangle\,\tau_{\rm m}^{\rm mag}} \, ,
\label{eq:eta-A-g}
\end{equation}
\begin{equation}
\eta_{\rm A}^{(q)} = \frac{62\zeta(6)N_c^2N_f}{\pi^2}
   \frac{T^6}{g^2\langle{\cal B}^2 \rangle\,\tau_{\rm m}^{\rm mag}} \, .
\label{eq:eta-A-q}
\end{equation}
\end{subequations}
These results differ by a numerical factor from the one obtained previously 
\cite{Asakawa:2006tc} by approximating $\bar\Delta(\mathbf{p})$ as a constant. 
We emphasize again that, given an ensemble of color fields, gluons and each 
flavor of quarks generate their separate contribution to the shear viscosity. 

The results obtained for Scenario B have a similar, but somewhat more 
complicated form, because the algebraic equation for $\bar\Delta(p)$
is replaced with a second-order differential equation. We will not discuss
this case further here and come back to it in the context of the calculation
of the complete shear viscosity.

However, we note that the field ensemble is sensitive to the total value 
of $\eta_{\rm A}$, because the field instabilities are driven by the 
overall anisotropy of the parton distribution. Possessing a smaller color 
charge than gluons, quarks develop a larger momentum space anisotropy in 
an expanding quark-gluon plasma, neglecting the effect of collisions.  On 
the other hand, quarks contribute with a smaller weight than gluons to the 
color polarization tensor, which drives the instability of soft modes of the 
color field. Thus, although $\bar\Delta(\mathbf{p})$ and $\eta_{\rm A}$ 
are inversely proportional to $C_2$, the contribution of gluons and quarks 
to the color field instabilities, which is proportional to $C_2\eta_{\rm A}$, 
is independent of the magnitude of their color charge.

\subsection{Complete Shear Viscosity}

The full linearized diffusive Vlasov-Boltzmann equation (\ref{eq:DVBE}) 
constitutes a linear integral equation for the scalar function 
$\bar\Delta(p)$ characterizing the deviation of the momentum
distribution from equilibrium. An exact solution of this equation 
requires numerical methods. The standard approach \cite{Landau10,deGroot}
makes use of the fact that the kernel of the linearized collision operator 
is self-adjoint with respect to an appropriately chosen scalar product 
and has non-negative eigenvalues. As a consequence, the solution of the
Boltzmann equation coincides with the minimum of a quadratic functional
$W[\bar{f}_1]$ and can thus be obtained from a variational principle.
The variational principle can be cast into the form:
\begin{equation}
W[\bar{f}_1] \equiv \int\frac{d^3p}{(2\pi)^3} \bar{f}_1({\mathbf p})
    \left[ v^\mu\frac{\partial f_0({\mathbf p})}{\partial x^\mu}
    + \frac{1}{2}\left(-\nabla_p\cdot D\cdot\nabla_p\delta\bar{f}({\mathbf p})
      + I[\bar{f}_1]({\mathbf p}) \right) \right]
= {\rm min.} 
\label{eq:var}
\end{equation}
The minimum of (\ref{eq:var}) defines the optimal solution $f_1(\mathbf{p})$
or $\bar\Delta(\mathbf{p})$ of the transport equation (\ref{eq:DVBE}).

The optimal function $\bar\Delta(p)$ can be determined by means 
of the variational method after an expansion in a complete set of orthogonal 
functions. A good (up to a few percent) approximation can be obtained by 
choosing the one-parameter function $\bar\Delta(p)=A|{\mathbf p}|/T$,
where $A$ is a constant. Taking the appropriate moment of the linearized 
transport equation results in an algebraic equation for $A$, which can be 
solved to obtain an approximate analytic expression for the shear viscosity. 
Because the mean free paths of quarks and gluons are different, we need to 
introduce different parameters for quarks ($A_q$) and gluons ($A_g$).

We now evaluate the momentum integrals in (\ref{eq:var}) for massless 
quarks and gluons. For the drift term we obtain:
\begin{eqnarray}
W_{\rm D}[\bar{f}_1] &=& \int\frac{d^3p}{(2\pi)^3} \bar{f}_1({\mathbf p})
    v^\mu\frac{\partial}{\partial x^\mu}f_0(p)
\nonumber \\
&=& -\frac{1}{T^3} \int \frac{d^3p}{(2\pi)^3E_p^2}\,
     \bar\Delta(p) \left[p_ip_j(\nabla u)_{ij}\right]^2 
     f_0(p)(1\pm f_0(p)) {\mathbf v}\cdot\nabla_r f_0(p)
\nonumber \\ 
&=& -\frac{2|\nabla u|^2}{9\pi^2T^3} 
  \int_0^\infty dp\,p^4 A f_0(p)
\nonumber \\ 
&=& -\frac{32|\nabla u|^2}{3\pi^2} \zeta(5) T^2 
     \left[(N_c^2-1)A_g + \frac{15}{8}N_cN_f A_q\right] .
\label{eq:mom-drift}
\end{eqnarray}
The diffusive Vlasov term yields (in Scenario A):
\begin{eqnarray}
W_{\rm V}^{\rm (A)}[\bar{f}_1] 
&=& -\frac{1}{2} \int\frac{d^3p}{(2\pi)^3} 
    \bar{f}_1({\mathbf p})\nabla_p\cdot D\cdot\nabla_p
    \delta\bar{f}({\mathbf p})
\nonumber \\
&=& \frac{3C_2g^2 \langle{\cal B}^2 \rangle\, \tau_{\rm m}^{\rm mag}}
         {2T^4(N_c^2-1)} 
    \int \frac{d^3p}{(2\pi)^3E_p^4}
    \bar\Delta(p)^2 \left[p_ip_j(\nabla u)_{ij}\right]^2 
    f_0(p)(1\pm f_0(p)) 
\nonumber \\ 
&=& \frac{4|\nabla u|^2}{15T^5} 
  \frac{C_2 g^2 \langle{\cal B}^2 \rangle\, \tau_{\rm m}^{\rm mag}}
       {N_c^2-1} 
  \int_0^\infty dp\,p^3 A^2 f_0(p)
\nonumber \\ 
&=& \frac{16|\nabla u|^2}{5\pi^2T} \zeta(4) 
    g^2\langle{\cal B}^2\rangle \tau_{\rm m}^{\rm mag}
    \left[ N_c A_g^2 + \frac{7}{8}N_f A_q^2 \right] .
\label{eq:mom-Vlasov-A}
\end{eqnarray}
For Scenario B, the diffusive Vlasov term receives contributions from
color-electric as well as color-magnetic fields:
\begin{eqnarray}
W_{\rm V}^{\rm (B)}[\bar{f}_1] 
&=& -\frac{1}{2} \int\frac{d^3p}{(2\pi)^3} 
    \bar{f}_1({\mathbf p})\nabla_p\cdot D\cdot\nabla_p
    \delta\bar{f}({\mathbf p})
\nonumber \\
&=& \frac{C_2g^2}{T^4(N_c^2-1)} 
    \int \frac{d^3p}{(2\pi)^3E_p^2}
    \left(\frac{\langle{\cal E}^2 \rangle\, \tau_{\rm m}^{\rm el}}{p^2}
         +\frac{\langle{\cal B}^2 \rangle\, \tau_{\rm m}^{\rm mag}}{E_p^2}
    \right)
\nonumber \\ 
&& \qquad\qquad\qquad
    \times \bar\Delta(p)^2 \left[p_ip_j(\nabla u)_{ij}\right]^2 
    f_0(p)(1\pm f_0(p)) 
\nonumber \\
&& - \frac{C_2g^2 \langle{\cal E}^2 \rangle\, \tau_{\rm m}^{\rm el}}
          {6T^4(N_c^2-1)} 
    \int \frac{d^3p}{(2\pi)^3p^3E_p}
    \bar\Delta(p) \left[p_ip_j(\nabla u)_{ij}\right]^2 
\nonumber \\ 
&& \qquad\qquad\qquad
    \times \frac{\partial^2}{\partial p^2}
    \left[\frac{p^3\bar\Delta(p)}{E_p}f_0(p)(1\pm f_0(p))\right]
\nonumber \\ 
&=& \frac{8|\nabla u|^2 C_2 g^2}{45T^5(N_c^2-1)} 
    (2\langle{\cal E}^2 \rangle\, \tau_{\rm m}^{\rm el}
     +\langle{\cal B}^2 \rangle\, \tau_{\rm m}^{\rm mag})
    \int_0^\infty dp\,p^3 A^2 f_0(p)
\nonumber \\ 
&=& \frac{32|\nabla u|^2}{15\pi^2T} \zeta(4) 
    g^2(2\langle{\cal E}^2\rangle \tau_{\rm m}^{\rm el}
        +\langle{\cal B}^2\rangle \tau_{\rm m}^{\rm mag})
    \left[ N_c A_g^2 + \frac{7}{8}N_f A_q^2 \right] .
\label{eq:mom-Vlasov-B}
\end{eqnarray}
This result differs from the one obtained for Scenario A only by 
the substitution
\begin{equation}
\langle{\cal B}^2\rangle \tau_{\rm m}^{\rm mag} 
\longrightarrow
\frac{2}{3}(2\langle{\cal E}^2\rangle \tau_{\rm m}^{\rm el}
            +\langle{\cal B}^2\rangle \tau_{\rm m}^{\rm mag}) \, .
\label{eq:AtoB}
\end{equation}

Finally, we simply state the result for the collision term in the 
leading logarithmic approximation and refer to Arnold {\em et 
al.}~\cite{Arnold:2000dr} for details of the calculation:
\begin{eqnarray}
W_{\rm C}[\bar{f}_1] &=& \frac{1}{2} \int\frac{d^3p}{(2\pi)^3} 
    \bar{f}_1({\mathbf p}) I[\bar{f}_1]({\mathbf p}) 
\nonumber \\
&=& \frac{\pi|\nabla u|^2T^2}{45} (N_c^2-1) g^4\ln g^{-1} 
    \left[ \frac{1}{9}(2N_c+N_f)
           \left(N_c A_g^2+\frac{7}{8}N_f A_q^2\right)
\right. \nonumber \\ 
& & \qquad\qquad\qquad \left.
    + \frac{\pi^2N_f}{256N_c}(N_c^2-1)(A_g-A_q)^2 \right] \, .
\label{eq:mom-coll}
\end{eqnarray}
Note that our expression differs from that of Ref.~\citen{Arnold:2000dr} 
by an overall factor $4|\nabla u|^2/(45T)$ owing to the different 
definition of the expectation value (\ref{eq:var}).

In order to calculate the shear viscosity, we have to minimize 
(\ref{eq:var}) with respect to $A_g$ and $A_q$, and then insert the 
obtained values into the expression 
\begin{equation}
\eta = \frac{24\zeta(5)T^3}{3\pi^2}
       \left[ (N_c^2-1)A_g + \frac{15}{8}N_cN_f A_q \right] 
\label{eq:eta-Agq}
\end{equation}
obtained by performing the momentum integral in the formula 
(\ref{eq:eta-Delta}) for the shear viscosity. The mininization results 
in two linear equations for $A_g$ and $A_q$. We state the expression for 
Scenario A: 
\begin{eqnarray}
\frac{32\zeta(5)}{3\pi^2}
\left(
\begin{array}{c}
N_c^2-1 \\
\frac{15}{8}N_cN_f
\end{array}
\right)
&=&
\frac{32\zeta(4)}{5\pi^2}
\frac{g^2 \langle{\cal B}^2 \rangle\, \tau_{\rm m}^{\rm mag}}{T^3}
\left(
\begin{array}{c}
N_c A_g \\
\frac{7}{8}N_f A_q
\end{array}
\right)
\nonumber \\
& & 
+ \frac{\pi(N_c^2-1)}{45}g^4\ln g^{-1} 
\left[
\frac{2}{9}(2N_c+N_f)
\left(
\begin{array}{c}
N_c A_g \\
\frac{7}{8}N_f A_q
\end{array}
\right) \right.
\nonumber  \\
&& \qquad\qquad \left. 
+ \frac{\pi^2N_f(N_c^2-1)}{128N_c} 
\left(
\begin{array}{c}
A_g-A_q \\
A_q-A_g 
\end{array}
\right)
\right] \, .
\label{eq:A-array}
\end{eqnarray}
In the absence of turbulent fields, this array of equations reduces 
to eqs.~(6.8), (6.9) of Ref.~\citen{Arnold:2000dr}. The result for 
Scenario B is obtained by means of the substitution (\ref{eq:AtoB}).

It is useful to introduce a concise vector notation for these equations.
With the help of the two-vector $A=(A_g,A_q)$, (\ref{eq:A-array}) 
can be written in the symbolic form
\begin{equation} 
(a_{\rm A} + a_{\rm C}) A = r \, ,
\label{eq:A-symb}
\end{equation}
with
\begin{subequations}
\label{eq:raA}
\begin{eqnarray}
r &=& \frac{32\zeta(5)}{3\pi^2}
\left(
\begin{array}{c}
N_c^2-1 \\
\frac{15}{8}N_cN_f
\end{array}
\right) \, ,
\\
a_{\rm A} &=& \frac{32\zeta(4)}{5\pi^2}
\frac{g^2 \langle{\cal B}^2 \rangle\, \tau_{\rm m}^{\rm mag}}{T^3}
\left(
\begin{array}{cc}
N_c & 0 \\
0 & \frac{7}{8}N_f
\end{array}
\right) \, ,
\\
a_{\rm C} &=& \frac{\pi(N_c^2-1)}{45}g^4\ln g^{-1} 
\left[
\frac{2}{9}(2N_c+N_f)
\left(
\begin{array}{cc}
N_c & 0 \\
0 & \frac{7}{8}N_f 
\end{array}
\right) \right.
\nonumber \\
&& \qquad\qquad \left.
+ \frac{\pi^2N_f(N_c^2-1)}{128N_c} 
\left(
\begin{array}{cc}
1 & -1 \\
-1 & 1  
\end{array}
\right)
\right] \, .
\end{eqnarray}
\end{subequations}
The equation (\ref{eq:eta-Agq}) for the shear viscosity then takes 
the form of a scalar product between the vectors $r$ and $A$:
\begin{equation}
\eta = \frac{3}{4} r\cdot A 
= \frac{3}{4}\, r\cdot (a_{\rm A}+a_{\rm C})^{-1}\cdot r \, ,
\label{eq:eta-rA}
\end{equation}
where we used (\ref{eq:A-symb}) to express $A$ in terms of $r$ and
the matrices $a_{\rm A}$ and $a_{\rm C}$. 

The inversely additive property of the contributions of turbulent color 
fields and parton collisions to the total shear viscosity is apparent 
in (\ref{eq:eta-rA}). Its origin lies in the additivity of the relaxation 
rates due to different processes and allows us to write the total shear 
viscosity as an inverse sum of anomalous and collisional viscosity:
\begin{equation}
\eta^{-1} \approx \eta_{\rm A}^{-1} + \eta_{\rm C}^{-1} \, ,
\label{eq:eta-total}
\end{equation}
with
\begin{eqnarray}
\eta_{\rm A} = \frac{3}{4}\, r\cdot a_{\rm A}^{-1}\cdot r \, ,
& \qquad &
\eta_{\rm C} = \frac{3}{4}\, r\cdot a_{\rm C}^{-1}\cdot r \, .
\label{eq:eta-AC}
\end{eqnarray}
Equation (\ref{eq:eta-total}) implies that, whichever contribution to 
$\eta$ is {\rm smaller}, dominates the overall shear viscosity. The 
anomalous viscosity dominates when $\eta_{\rm A}<\eta_{\rm C}$. Because 
$\eta_{\rm A}$ grows with a smaller power of $g$ than $\eta_{\rm C}$,
the anomalous viscosity dominates for sufficiently weak coupling, 
and we thus have $\eta \approx \eta_{\rm A}$ for $g\to 0$.

\subsection{Estimate of the anomalous viscosity}

Equation (\ref{eq:eta-A1}) shows that $\eta_{\rm A}$ decreases with 
increasing strength of the turbulent fields. Since the amplitude of 
these fields grows with the magnitude of the momentum anisotropy, a 
large anisotropy will result in a small value of $\eta_{\rm A}$. 
The anomalous mechanism thus exhibits a stable equilibrium in which 
the shear viscosity regulates itself: The momentum anisotropy grows 
with $\eta_{\rm A}$, but an increased anisotropy tends to reduce the 
anomalous viscosity. This leads to a self-consistency condition which 
determines $\eta_{\rm A}$.

In order to proceed further, we need to explore the dependence of the
turbulent fields on the anisotropy of the momentum distribution of the 
partons. In particular, we must know how large $\langle{\cal B}^2 \rangle$
and $\tau_{\rm m}$ are. The coherent color magnetic fields are only
generated by the plasma instability when the momentum distribution 
of partons in the quark-gluon plasma is deformed due to the collective
expansion. Analytical studies have shown that the instability always
occurs when the momentum distribution is anisotropic. We also know 
from these studies how the growth rate of the instability depends on
the anisotropy of the momentum distribution, but there are no published
systematic studies that show how the ``saturation'' level of the coherent 
field energy depends on the anisotropy. We will therefore rely on some
heuristic arguments for the needed dependences.

The study by Romatschke and Strickland \cite{Romatschke:2003ms} uses 
the following parametrization of the anisotropic momentum distribution:
\begin{equation}
f({\mathbf p}) = f_0\left(\sqrt{p^2+\xi({\mathbf p}\cdot\hat{n})^2}\right) 
\approx f_0(p) - \frac{\xi({\mathbf p}\cdot\hat{n})^2}{2E_pT} f_0(1\pm f_0) \, .
\label{eq:RS}
\end{equation}
Choosing $\hat{n}=\hat{e}_z$ and subtracting the trace, this corresponds 
to a perturbation of the equilibrium distribution:
\begin{equation}
f_1({\mathbf p}) = - \frac{\xi}{2E_pT} \left(p_z^2 - \frac{p^2}{3}\right) \, .
\label{eq:f1-RS}
\end{equation}
Comparing with (\ref{eq:f1-Qzz}) this establishes the connection
\begin{equation}
\xi = 2\bar\Delta\, \frac{|\nabla u|}{T} .
\label{eq:xi-Delta}
\end{equation}
The relative anisotropy of the stress tensor is given by
\begin{equation}
2\frac{\delta T_{xx}}{T_{xx}^{(0)}} 
= 2\frac{\delta T_{yy}}{T_{yy}^{(0)}} 
= -\frac{\delta T_{zz}}{T_{zz}^{(0)}} 
= \frac{8}{15} \xi .
\label{eq:T1zz-T0zz}
\end{equation}
Comparing with (\ref{eq:Tzz}) we obtain a relation between $\eta$
and $\xi$, which takes the form (for a massless parton gas):
\begin{equation}
\xi = \frac{15\eta |\nabla u|}{2T_{00}}
= 10 \frac{\eta}{s}\, \frac{|\nabla u|}{T} \, .
\label{eq:xi-eta}
\end{equation}

The central point of our argument is that the average collective 
color field energy is a function of the momentum anisotropy. For 
Scenario A (turbulent color-magnetic fields only) this means:
$\langle{\cal B}^2\rangle = b(\xi)$. We do not know this function
in detail, but we know that $b(0)=0$, because no instability exists 
in the absence of a momentum anisotropy. The simplest {\em ansatz} is
a power law, which we will write in the form
\begin{equation}
g^2 \langle{\cal B}^2 \rangle = b_0 g^4 T^4 \xi^n 
\label{eq:B-sat}
\end{equation}
with an as yet unknown power $n$. 
Following the discussion before equation (\ref{eq:gB2r}) we set 
$n=2$.\footnote{This differs from the assumption made in 
Ref.~\citen{Asakawa:2006tc}, where we assumed $n=1$.} The memory time 
$\tau_{\rm m}$ can be determined by one of two mechanisms. If the plasma 
particles move faster than the coherent fields propagate, $\tau_{\rm m}$ 
will be set by the spatial coherence length of the coherent fields.
This coherence length is given by the wave length of the maximally
unstable mode, which is of the order of the Debye length:\footnote{In 
the opposite situation, when the color fields evolve rapidly
compared with the motion of the plasma particles, the memory time
is determined by the decoherence time of the fields interacting 
with the particles and among themselves via the nonlinearities 
of the Yang-Mills equations. The case, in which the back reaction 
of the particle motion on the unstable field modes determines the
decoherence time has been treated by Dupree \cite{Dupree:1966} 
and by Abe and Niu.\cite{Abe:1980a,Abe:1980b}}
\begin{equation}
\tau_{\rm m} = d_0 \xi^{-1/2}(gT)^{-1} .
\label{eq:taumem}
\end{equation}
The decoherence time and its dependence on $\xi$ can, in principle, 
be determined from simulations of the classical Yang-Mills equations. 
In the absence of such a determination, we shall assume that 
$\tau_{\rm m}$ is given by (\ref{eq:taumem}).

We are now ready to derive the self-consistency condition for the 
anomalous viscosity in the limit when the collisional viscosity can
be neglected. We are interested in the late time, steady state 
situation where the expansion drives the momentum distribution into a 
slightly oblate anisotropy along the $z$-axis. As discussed before, 
this leads to the growth of unstable mean field modes, which determines 
the viscosity $\eta_{\rm A}$ which, in turn, controls the size of the 
momentum space anisotropy. The feedback loop is a stable one, because 
a large anisotropy leads to large saturation levels of the fields, 
which reduce the viscosity and thus limit the size of the momentum 
anisotropy. 

One technical complication of the limit of vanishing collisional 
viscosity is that the momentum distributions of quarks and gluons 
then attain different anisotropies, because color charges differ
and thus their interaction with the turbulent fields is different.
This implies that the presently unknown constants $b_0$ and $d_0$ 
receive different contributions from quarks and gluons, and that 
we should, in principle, distinguish between $\xi_g$ and $\xi_q$.
In order to avoid unnecessay distractions from our argument, and 
because the numerical constants are not known anyway, we will only
consider the gluon contribution to the anomalous viscosity in the 
following. The generalization to include quarks is straightforward 
and does not change the functional dependence of the result on $g$,
$T$, and $|\nabla u|$.

Combining (\ref{eq:B-sat}) and (\ref{eq:taumem}) into a single scaling
relation and replacing $\xi$ with $\eta=\eta_{\rm A}$, we obtain:
\begin{equation}
g^2\langle{\cal B}^2 \rangle\, \tau_{\rm m} 
= b_0d_0 (gT)^3 \xi^{3/2}
= b_0d_0 (gT)^3 \left(\frac{10\eta_{\rm A}|\nabla u|}{sT}\right)^{3/2} \, .
\label{eq:g2B2t}
\end{equation}
We now insert this result into the expressions (\ref{eq:eta-A-g}) for 
the anomalous viscosity and obtain the self-consistency relation:
\begin{equation}
\frac{\eta_{\rm A}}{s} 
= c_0 \left(\frac{Ts}{g^2|\nabla u|\eta_{\rm A}}\right)^{3/2} \, ,
\label{eq:eta-eta}
\end{equation}
where we have combined all numerical constants into a single factor $c_0$. 
We note once again that this relation neglects contributions from the
collisional shear viscosity $\eta_{\rm C}$ and does not distinguish between
the contributions from quarks and gluons. We also note that the power on
the right-hand side differs slightly from that in Ref.~\citen{Asakawa:2006tc},
because of our different scaling assumptions in (\ref{eq:B-sat}) and 
(\ref{eq:taumem}). Resolving (\ref{eq:eta-eta}) for $\eta_{\rm A}/s$ 
finally yields the desired expression for the self-consistent anomalous
shear viscosity:
\begin{equation}
\frac{\eta_{\rm A}}{s} = 
\bar{c}_0 \left(\frac{T}{g^2|\nabla u|}\right)^{3/5} \, .
\label{eq:eta-A}
\end{equation}
\begin{figure}[t]
\centerline{\includegraphics[width=0.8\linewidth]{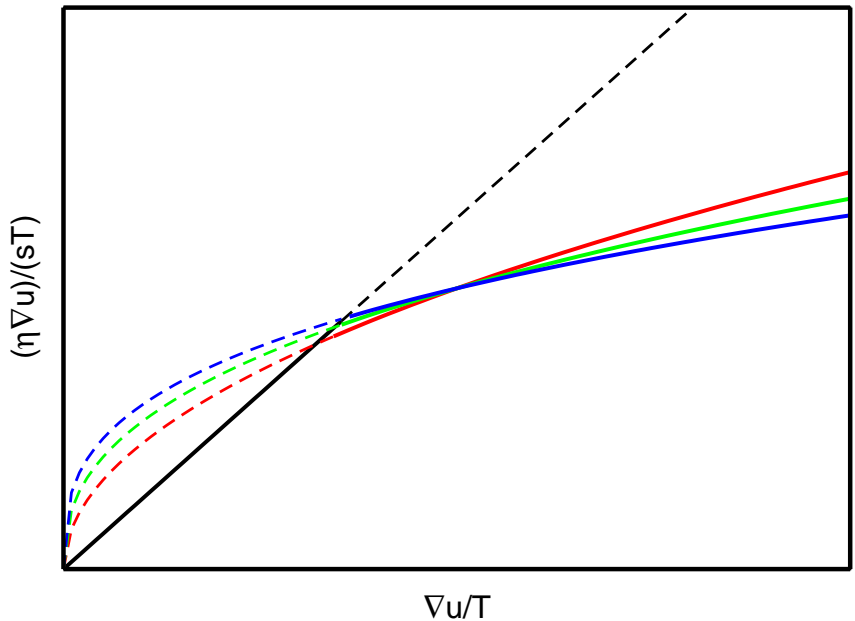}}
\caption{Schematic representation of the dependence of the collisional
and anomalous viscous stress $\delta T_{ik}$ on the velocity gradient.
The collisional viscous stress is shown by the linearly rising black 
line; the anomalous viscous stress is shown by the (colored) curved 
lines for three scaling exponents of the turbulent color field energy 
($n=1.5,2,2.5$). Solid lines indicate the dominant source of viscous 
stress in different regions of the scaled velocity gradient $|\nabla u|/T$.}
\label{fig1}
\end{figure}

Several things are noteworthy about this result. First, if the memory
time $\tau_{\rm m}$ is longer than our estimate (\ref{eq:taumem}),
the value of $\eta_{\rm A}$ decreases. Second, the dependence on the 
gauge coupling ($\sim g^{-6/5}$) is parametrically much weaker than that
of the collisional viscosity ($\sim (g^4\ln(1/g))^{-1}$). Thus, for early 
times $\tau$ and weak coupling $g\ll 1$, the anomalous viscosity will be 
much smaller than the collisional viscosity and thus dominate the 
total viscosity according to (\ref{eq:eta-total}). 

It is instructive to consider the role of the anomalous viscosity in 
the hydrodynamical equations. As (\ref{eq:Tzz}) shows, the viscous 
contribution to the stress tensor is proportional to $\eta |\nabla u|$.
For the collisional shear viscosity which is, in first approximation, 
independent of the magnitude of the velocity gradient; this implies 
that $\delta T_{ik}$ grows linearly with $|\nabla u|$. The anomalous
shear viscosity (\ref{eq:eta-A}), on the other hand, is a decreasing 
function of the velocity gradient; its contribution to the stress 
tensor grows like $|\nabla u|^{2/5}$ for our scaling assumptions.
The unusual dependence of $\eta_{\rm A}$ on $|\nabla u|$ certainly 
justifies the term ``anomalous viscosity''.

The different dependence of the collisional and the anomalous viscous
stress tensor on the velocity gradient is shown schematically in 
Fig.~\ref{fig1}. For very small gradients the linear dependence of 
the collisional viscous stress tensor dominates, but for larger 
velocity gradients the lower power associated with the anomalous 
shear viscosity will assert its dominance. The precise location of 
the cross-over between the two domains depends on the value of the
numerical constant $\bar{c}_0$, but we can deduce from (\ref{eq:eta-A})
that the cross-over point shifts to lower values of $|\nabla u|$ with
decreasing coupling constant $g$. We also show in the figure the 
effect of choosing a different power $n$ in the scaling law 
(\ref{eq:B-sat}) for the energy density of the turbulent color 
fields. Finally, we note that the decreasing dependence of $\eta_{\rm A}$
on $|\nabla u|$ implies that the shear viscosity can, in principle, fall 
below the Kovtun-Son-Starinets (KSS) bound $\eta/s = (4\pi)^{-1}$.

\section{Conclusions and Outlook}

In this paper we have presented details of our derivation of the anomalous
viscosity in an anisotropically expanding quark-gluon plasma, which arises 
from interactions of thermal partons with dynamically generated color fields.  
In the weak coupling limit or for large velocity gradients, the anomalous 
viscosity is much smaller than the viscosity due to collisions among thermal 
partons. By reducing the shear viscosity of a weakly coupled, but expanding
quark-gluon plasma, this mechanism could possibly explain the observations 
of the RHIC experiments without the assumption of a strongly coupled plasma 
state. 

Due to the self-consistency condition, the anomalous shear viscosity 
itself is inversely dependent on the expansion rate of the plasma. 
This means that the viscous term in the hydrodynamic equation
does not depend on the velocity gradient linearly, but sublinearly.
This unusual dependence on the velocity gradient justifies the term
``anomalous'' viscosity. It also implies that the response of the medium
to the expansion is nonlinear, in fact less than linear. The usual method
to relate the transport coefficient to a correlation function (Kubo formula)
that can be measured in full equilibrium is therefore not applicable. Even 
if the appropriate real-time correlation function could be determined from
Euclidean lattice QCD simulations, it would fail to describe the anomalous
viscosity. More sophisticated methods, including a treatment of the color
instabilities in the expanding quark-gluon plasma, will be necessary.

The presence of anomalous viscosity for a rapidly expanding 
{\em anisotropic} quark-gluon plasma has important consequences for the 
early Universe. The expansion of the quark-gluon plasma in the early 
Universe is relatively slow and thermalization is therefore always 
maintained. Thus, the quark-gluon plasma in the early Universe does not
generally possess an anomalous shear viscosity. However, when 
off-equilibrium processes induce a strong local anisotropy, the 
anomalous viscosity can become important in determining the total 
viscosity. This could have been the case during the reheating period 
after inflation and the electroweak phase transition, and may have 
affected the production of fluctuations in the Universe, 
baryogenesis, etc.

The approach described here can be applied to other transport properties of 
an expanding, turbulent quark-gluon plasma. Examples are the coefficient 
$\hat{q}$ of radiative energy loss of an energetic parton,\cite{Baier:2000mf} \
the absorption coefficient for bound states of heavy quarks (e.~g.\ J/$\psi$), 
and flavor equilibration rates. For the rate of strange quark pair production, 
we note that a quark pair can be produced in the presence of a mean color 
field by a single gluon. This process is well known in quantum electrodynamics,
where a single photon can produce an electron-positron pair in the presence
of a strong electromagnetic field, such as the Coulomb field of a nucleus.
In QED the pair production rate grows like $Z^2$, where $Ze$ is the nuclear
charge; in the case of the quark-gluon plasma one would expect the rate to
grow like $g^2\langle{\cal B}^2\rangle$ and hence be proportional to the 
expansion rate. A rapid expansion would, therefore, accelerate the rate 
of light quark pair production. Such an effect may, indeed, contribute to
the very high rate of quark pair production seen in numerical solutions 
of the Dirac equation in longitudinally expanding gluon fields in the
color glass condensate picture.\cite{Gelis:2005pb}

Turbulent color fields can also contribute to the dissociation of 
heavy quark-antiquark bound states, because the two constituents will 
be deflected in opposite directions. This effect has been considered for
randomly oriented color-electric fields,\cite{Hufner:1989fn} \ but not
for color-magnetic fields, which should have a similar effect, albeit
suppressed by a factor $(v/c)^2$, where $v$ is the velocity of the 
charmonium. Turbulent color fields will also contribute to the diffusion
coefficient for heavy quarks in the plasma. Finally, turbulent color 
fields can influence the trajectories of the partons contained in a jet 
created by a hard scattered quark or gluon. Because the fields created 
by the color instabilities in an expanding medium are polarized transversely 
to the expansion direction, the color magnetic fields will preferentially 
deflect the outgoing partons in the longitudinal direction and thus cause 
a longitudinal broadening of the jet cone. Such an effect has been been 
observed at RHIC.\cite{Wang:2004kf}

In a forthcoming publication we shall further investigate the phenomenological
consequences of the anomalous viscosity: in particular we shall focus
on jet energy-loss in turbulent color fields and address the question whether
the vast array of current data at RHIC can be understood in the framework of a
weakly coupled QGP with anomalous transport properties.

\section*{Acknowledgements}
This work was supported in part by grants from the U.~S.~Department 
of Energy (DE-FG02-05ER41367), the National Science Foundation 
(NSF-INT-03-35392), and the Japanese Ministry of Education 
(Grant-in Aid No. 17540255). BM thanks the members of the Yukawa
Institute for Theoretical Physics, especially T.~Kunihiro, for 
enlightening discussions and hospitality during his stay in Kyoto.
We thank J.~Ruppert for valuable comments on the manuscript.

\appendix
\section{Alternative Derivation of the Vlasov Diffusion Term}

In this section, we present a different derivation of the Vlasov 
diffusion term (\ref{eq:VT-2}), based on the evolution of the 
extended phase-space distribution ${\tilde f}(r,p,Q,t)$,
which is given by \cite{Heinz:1983nx}
\begin{equation}
\left[ v^\mu \left( \frac{\partial}{\partial x^\mu}
  + g f_{abc} Q^a A^b_\mu \frac{\partial}{\partial Q^c} \right) 
  + g Q^a \left({\cal E}^a + {\mathbf v}\times{\cal B}^a\right)
  \cdot\nabla_p \right] {\tilde f}(r,p,Q,t) = 0 .
\label{eq:Vlasov-Q}
\end{equation}
Here we have suppressed any unnecessary vector notation and omitted the
collision term. Our alternative derivation of the diffusion term starts 
from the integral representation for ${\tilde f}$:
\begin{equation}
{\tilde f}(r,p,Q,t) = \int dr_0\,dp_0\,dQ_0\, \delta(r-\bar{r}(t))\, 
  \delta(p-\bar{p}(t))\, \delta(Q-\bar{Q}(t))\, {\tilde f}(r_0,p_0,Q_0,0) ,
\label{eq:f-evol}
\end{equation}
with
\begin{subequations}
\label{eq:eom}
\begin{eqnarray}
\frac{d\bar{r}}{dt} &=& \bar{v} = \frac{\bar{p}}{E_p} \, ,
\\
\frac{d\bar{p}}{dt} &=& g \bar{Q}^a \left( {\cal E}^a(\bar{r}) + 
  \bar{v}\times {\cal B}^a(\bar{r}) \right) \, ,
\\
\frac{d\bar{Q}}{dt} &=& g f_{abc} \bar{Q}^b A^c_\mu v^\mu \, . 
\end{eqnarray}
\end{subequations}
and $r_0=\bar{r}(0)$, $\bar{p}(0)=p_0$, $\bar{Q}(0)=Q_0$. Here 
$\bar{r}(t)$ denotes the trajectory of a plasma particle, which is 
found at position $r_0$ with momentum $p_0$ and color charge $Q_0$ 
at time $t=0$. The evolving phase space distribution is given by the sum 
over all the trajectories of these (test) particles. When we act on 
(\ref{eq:f-evol}) with the Vlasov drift operator 
$(\partial_t + v\cdot\nabla_r)$ and use the definition (\ref{eq:eom}) 
of $\bar{v}$, we obtain:
\begin{eqnarray}
&& \left(\frac{\partial}{\partial t} + {\mathbf v}\cdot\nabla_r
  + g f_{abc} Q^a A^b_\mu v^\mu \frac{\partial}{\partial Q^c} \right) 
  {\tilde f}(r,p,Q,t)
\nonumber \\ 
&& \qquad\qquad\qquad
= - \int dr_0\,dp_0\, \delta(r-\bar{r})\,
    \frac{d\bar{{\mathbf p}}}{dt}\cdot\nabla_p\, 
    \delta(p-\bar{p})\, {\tilde f}(r_0,p_0,Q_0,0) ,
\label{eq:VE-1}
\end{eqnarray}
where we dropped the explicit notation of the time-dependence of $\bar{r}$,
$\bar{p}$, and $\bar{Q}$. Since $\bar{p}$ is time independent for vanishing 
color field, it makes sense to expand the delta function $\delta(p-\bar{p})$ 
around $(p-p_0)$ for weak fields (or short times). It is important here
to remember that (strong) plasma turbulence does not imply strong fields, 
just a spectral field distribution without phase correlations on all 
inverse length scales $k$. The expansion is:
\begin{equation}
\delta(p-\bar{p}) 
= \delta(p-p_0) - \Delta{\mathbf p}(t)\cdot\nabla_p \delta(p-p_0) + \cdots ,
\label{eq:delta-exp}
\end{equation}
where $\Delta p(t) = \bar{p}(t)-p_0$, implying $d(\Delta p)/dt=d\bar{p}/dt$.
The right-hand side of (\ref{eq:VE-1}) then becomes:
\begin{equation}
 - \int dr_0\,dp_0\, \delta(r-\bar{r})\,\frac{d\Delta{\mathbf p}}{dt}
   \cdot\nabla_p 
   \left[1-\Delta{\mathbf p}(t)\cdot\nabla_p\right] \delta(p-p_0) 
   {\tilde f}(r_0,p_0,Q_0,0) 
   + \cdots .
\label{eq:VE-2}
\end{equation}
The expression now contains a term quadratic in $\Delta p$, and it 
makes sense to take an ensemble average over the color fields. We also 
integrate by parts with respect to $\nabla_p$ and obtain:
\begin{eqnarray}
\left\langle \left(\frac{\partial}{\partial t}+{\mathbf v}\cdot\nabla_r
    + g f_{abc} Q^a A^b_\mu v^\mu \frac{\partial}{\partial Q^c}\right)
    {\tilde f} \right\rangle
&=& \left\langle \nabla_p\cdot\Delta{\mathbf p}(t)\frac{d\Delta{\mathbf p}}{dt}
    \cdot\nabla_p\, {\tilde f} \right\rangle
\nonumber \\ 
&\equiv & \nabla_p \cdot D \cdot\nabla_p \langle {\tilde f}\rangle .
\label{eq:VE-diff}
\end{eqnarray}
In the last step we have introduced the diffusion coefficient $D$ and
assumed that we can factorize the ensemble average. Explicitly, the 
diffusion coefficient is given by ($i,j=x,y,z$):
\begin{equation}
D_{ij} \equiv \left\langle \Delta p_i(t)\frac{d\Delta p_j}{dt}\right\rangle
= g^2 \int_0^t dt' \left\langle Q^a(t') F^a_i(\bar{r}(t'),t') 
  Q^b(t) F^b_j(\bar{r}(t),t) \right\rangle ,
\label{eq:D-def}
\end{equation}
where 
\begin{equation}
F^a_i(\bar{r}(t),t) = {\cal E}^a_i(\bar{r}(t),t) 
  + \left(\bar{v}(t)\times {\cal B}^a(\bar{r}(t),t)\right)_i .
\label{eq:field}
\end{equation}
In the local rest frame of the medium the electric and magnetic components 
of the color field can be expressed as (using the convention 
$\epsilon^{0123}=1$):
\begin{equation}
{\cal E}^a_i = F^a_{i\nu}u^\nu \, ; \qquad
{\cal B}^a_i = \epsilon_{i\lambda\mu\nu} F^{a\lambda\mu}u^\nu \, .
\label{eq:field-u}
\end{equation}
If we now argue that the correlation time/length for the color fields is 
short in comparison with the temporal change of the velocity of a plasma 
particle, we can take $\bar{v}(t)=\bar{v}(t')=v$ out of the average and 
are left with the autocorrelation function of the color fields and color 
charges along a typical particle trajectory.

The time evolution of the color charge $Q^a(t)$ is given by the solution 
of the last equation (\ref{eq:eom}):
\begin{equation}
Q^a(t') = P\, \exp\left( \int_{\bar{r}(t)}^{\bar{r}(t')} f_{abc}
  A^b_\mu\, dx^\mu \right) Q^c(t)
= U_{ac}(\bar{r}(t'),\bar{r}(t)) Q^c(t) \, .
\label{eq:Q-evol}
\end{equation}
Inserting this solution into (\ref{eq:D-def}), we obtain
\begin{eqnarray}
D_{ij} & \equiv & 
   \left\langle \Delta p_i(t)\frac{d\Delta p_j}{dt}\right\rangle
\nonumber \\
&=& g^2 \int_0^t dt' \left\langle Q^c(t) Q^b(t) F^a_i(\bar{r}(t'),t') 
    U_{ac}(\bar{r}(t'),\bar{r}(t)) F^b_j(\bar{r}(t),t) \right\rangle \, .
\label{eq:D-def2}
\end{eqnarray}
If we now assume that the distribution of the color charges of partons 
at a given time $t$ is random and independent of the color fields, 
$\langle Q^a Q^b \rangle = (N_c^2-1)^{-1}C_2\delta_{ab}$, we recover 
the expression (\ref{eq:VT-2}) for the Vlasov diffusion term with the 
field correlators (\ref{eq:F-correl}).

Finally, in order to make contact with the diffusive transport equation
(\ref{eq:DVBE}), we need to integrate (\ref{eq:VE-diff}) over the color
charges $Q$ to obtain an equation for 
\begin{equation}
\bar{f}(r,p,t) = \int dQ \langle{\tilde f}(r,p,Q,t)\rangle \, .
\label{eq:f-fQ}
\end{equation}
Partial integration with respect to $Q$ shows that the third term on 
the left-hand side of (\ref{eq:VE-diff}) vanishes in view of the
relation $\partial Q^a/\partial Q^c = \delta_{ac}$ and the antisymmetry 
of $f_{abc}$.

We end this section with a remark concerning abelian plasmas. Our 
derivation of the diffusive Vlasov term was motivated by the insight
that the color instabilities of an anisotropic nonabelian plasma 
saturate under the action of the nonlinearities of the Yang-Mills 
equation.\cite{Rebhan:2005re,Arnold:2005vb} \ 
This mechanism is, of course, absent in abelian plasmas. 
The saturation of the instablilities is then caused by the back 
reaction of the growing soft field modes on the particle 
distribution.\cite{Abe:1980a} \ In order to address this situation, we 
follow Dupree's argument \cite{Dupree:1966} that the ensemble average in
(\ref{eq:D-def}) should be taken over Fourier components of the color 
field, because these are the slowly varying variables. This is standard 
practice when dealing with an ensemble of waves. It certainly makes 
sense for electromagnetic plasmas, where nonrelativistic particles move 
in fields that propagate at the speed of light. In the case of the 
quark-gluon plasma, however, the situation is reversed: the thermal 
partons move with the speed of light, but the soft color fields 
obey a dispersion relation with a slower propagation speed. Thus it
is doubtful whether a similar reasoning would make sense. Nevertheless, 
for the sake of interest, we outline Dupree's approach (for 
color-magnetic fields only). One writes
\begin{equation}
{\cal B}^a(\bar{r}(t),t) = \sum_k {\cal B}^a(k)\, 
  e^{-i\omega_kt+ik\cdot\bar{r}(t)} ,
\label{eq:Bk}
\end{equation}
where ${\cal B}(k)$ are the Fourier components of the field. One can 
then pull the slow variables ${\cal B}(k)$ out of the time integral in 
(\ref{eq:D-def}). After factorizing the ensemble average one obtains
\begin{equation}
\nabla_p\cdot D \cdot \nabla_p
= - g^2 Q^a Q^b \sum_k \langle{\cal B}^a_i(k) {\cal B}^b_j(-k) \rangle 
    L^{(p)}_i L^{(p)}_j \left\langle
    \int_0^t dt' e^{-\omega_k(t-t')+ik\cdot(\bar{r}(t')-\bar{r}(t))}
    \right\rangle .
\label{eq:diff-D}
\end{equation}
The time integral can be interpreted as an autocorrelation or
memory time for the action of the magnetic fields on the particles.
In Dupree's approach, the value of the time integral is governed by 
the effect of the turbulent magnetic fields on the particle trajectory, 
and can be shown to satisfy a self-consistency condition.
We will not pursue this approach further here and refer the interested
reader to Abe and Niu's work.\cite{Abe:1980b} \

\section{Color Instabilities near Equilibrium}

For convenience, we here state the results for the growth rate of 
unstable quark-gluon plasma modes in a background distribution of 
quasi-thermal partons whose momentum distribution is only slightly
perturbed away from equilibrium. We follow the notation and derivation
of Romatschke and Strickland.\cite{Romatschke:2003ms} 

We assume that the momentum distribution in the rest frame of the 
medium can be written as
\begin{equation}
f({\mathbf p})
=f_0\left(\sqrt{{\mathbf p}^2+\xi({\mathbf p}\cdot{\mathbf n})^2}\right) ,
\label{eq:A1}
\end{equation}
where ${\mathbf n}$ is a unit vector defining the orientation of the 
anisotropy and $\xi\ll 1$. We denote the angle between the wave vector
${\mathbf k}$ of the considered field mode and ${\mathbf n}$ by $\theta$:
$\cos\theta = \hat{{\mathbf k}}\cdot{\mathbf n}$, where $ \hat{{\mathbf k}}$ 
is the unit vector in the direction of ${\mathbf k}$. For a given temperature 
$T$, the polarization function of the gluon field is then a function of the
variables $\omega$, $k=|{\mathbf k}|$, and $\theta$. Because the parton
distribution violates spherical symmetry, there are four different 
components of the polarization tensor, which can be expressed in terms
of the functions $\alpha,\beta,\gamma,\delta$. The gluon propagator in
medium can be decomposed with the help of the projectors $i,j=1,2,3$,
\begin{subequations}
\label{eq:A2}
\begin{eqnarray}
A_{ij} &=& \delta_{ij} - \hat{k}_i\hat{k}_j \, ,
\\
B_{ij} &=& \hat{k}_i\hat{k}_j \, ,
\\
C_{ij} &=& \tilde{n}_i\tilde{n}_j \, ,
\\
D_{ij} &=& \hat{k}_i\tilde{n}_j + \tilde{n}_i\hat{k}_j \, ,
\end{eqnarray}
\end{subequations}
where $\tilde{n} = {\mathbf n}-\hat{{\mathbf k}}(\hat{{\mathbf k}}
\cdot{\mathbf n})$. The resulting expression for the gluon propagator is
\begin{equation}
\Delta_{ij}(k,\omega,\theta) = \Delta_{\rm A}(A_{ij}-C_{ij}) 
+ \Delta_{\rm G}[(k^2-\omega^2+\alpha+\gamma)B_{ij}
  + (\beta-\omega^2)C_{ij} - \delta D_{ij}]
\label{eq:A3}
\end{equation}
with
\begin{subequations}
\label{eq:A4}
\begin{eqnarray}
\Delta_{\rm A}^{-1}(k,\omega,\theta) &=& k^2 - \omega^2 + \alpha ,
\\
\Delta_{\rm G}^{-1}(k,\omega,\theta) &=& (k^2 - \omega^2 + \alpha + \gamma)
  (\beta-\omega^2) - \tilde{\mathbf n}^2\delta^2  .
\end{eqnarray}
\end{subequations}
For small values of $\xi$ the functions $\alpha,\beta,\gamma,\delta$ are 
given by
\begin{subequations}
\label{eq:A5}
\begin{eqnarray}
\alpha &=& \Pi_{\rm T}(z) + \xi \left[
  \left(\frac{m_D^2}{3}-\Pi_{\rm T}(z)\right)\frac{z^2}{2}(5\cos^2\theta-1)
  - \frac{m_D^2}{3}\cos^2\theta \right.
\nonumber \\ && \qquad\qquad\qquad\qquad \left.
  + \frac{1}{2}\Pi_{\rm T}(z)(3\cos^2\theta-1) \right] ,
\\
\beta &=& z^2\Pi_{\rm L}(z) + \xi z^2\left[
  \left(\frac{m_D^2}{3}-z^2\Pi_{\rm L}(z)\right)(3\cos^2\theta-1) \right.
\nonumber \\ && \qquad\qquad\qquad\qquad \left.
  + \Pi_{\rm L}(z)(2\cos^2\theta-1)\right] ,
\\
\gamma &=& \xi \left(\Pi_{\rm T}(z)-\frac{m_D^2}{3}\right)(z^2-1)\sin^2\theta ,
\\
\delta &=& \xi \left(4z^2\frac{m_D^2}{3}-\Pi_{\rm T}(z)(1-4z^2)\right)
  \cos\theta ,
\end{eqnarray}
\end{subequations}
where $z=\omega/k$ and 
\begin{subequations}
\label{eq:A6}
\begin{eqnarray}
\Pi_{\rm T}(z) &=& \frac{m_D^2}{2}z^2\left[ 
  1-\left(\frac{z}{2}-\frac{1}{2z}\right)\ln\frac{z+1}{z-1}\right] ,
\\
\Pi_{\rm L}(z) &=& m_D^2 \left[\frac{z}{2}\ln\frac{z+1}{z-1} -1\right]
\end{eqnarray}
\end{subequations}
are the usual expressions for the transverse and longitudinal gluon 
polarization functions in thermal equilibrium.


\begin{thebibliography}{99}
  
\bibitem{RHIC_WP}
  I.~Arsene {\em et al.}, 
  \NPA{757,2005,1}.
\\
  B.~B.~Back {\em et al.},
  \NPA{757,2005,28}.
\\
  J.~Adams {\em et al.},
  \NPA{757,2005,102}.
\\
  K.~Adcox {\em et al.},
  \NPA{757,2005,184}.

\bibitem{Heinz:2001xi}
  U.~W.~Heinz and P.~F.~Kolb,
  \NPA{702,2002,269}.

\bibitem{Teaney:2003pb}
  D.~Teaney,
  \PRC{68,2003,034913}.

\bibitem{Kovtun:2004de}
  P.~Kovtun, D.~T.~Son and A.~O.~Starinets,
  \PRL{94,2005,111601}.

\bibitem{Asakawa:2006tc}
  M.~Asakawa, S.~A.~Bass and B.~M\"uller,
  \PRL{96,2006,252301}.

\bibitem{Arnold:2000dr}
  P.~Arnold, G.~D.~Moore and L.~G.~Yaffe,
  \JHEP{0011,2000,001}.

\bibitem{Mrowczynski:1988dz}
  S.~Mrowczynski,
  \PLB{214,1988,587}.

\bibitem{Mrowczynski:1993qm}
  S.~Mrowczynski,
  \PLB{314,1993,118}.

\bibitem{Romatschke:2003ms}
  P.~Romatschke and M.~Strickland,
  \PRD{68,2003,036004}.

\bibitem{Arnold:2005ef}
  P.~Arnold and G.~D.~Moore,
  \PRD{73,2006,025006}.

\bibitem{Arnold:2005qs}
  P.~Arnold and G.~D.~Moore,
  \PRD{73,2006,025013}.

\bibitem{Ludlam:2005gx}
  T.~Ludlam,
  \NPA{750,2005,9}.

\bibitem{Rafelski:2001hp}
  J.~Rafelski, J.~Letessier and G.~Torrieri,
  \PRC{64,2001,054907}.
  [Erratum \textbf{65} (2002), 069902.

\bibitem{Braun-Munzinger:2001ip}
  P.~Braun-Munzinger, D.~Magestro, K.~Redlich and J.~Stachel,
  \PLB{518,2001,41}.

\bibitem{Adams:2003xp}
  J.~Adams {\it et al.}  [STAR Collaboration],
  \PRL{92,2004,112301}.

\bibitem{Adcox:2003nr}
  K.~Adcox {\it et al.}  [PHENIX Collaboration],
  \PRC{69,2004,024904}.

\bibitem{Adler:2001nb}
  C.~Adler {\it et al.}  [STAR Collaboration],
  \PRL{87,2001,182301}.

\bibitem{Adler:2003kt}
  S.~S.~Adler {\it et al.}  [PHENIX Collaboration],
  \PRL{91,2003,182301}.

\bibitem{Adams:2003am}
  J.~Adams {\it et al.}  [STAR Collaboration],
  \PRL{92,2004,052302}.

\bibitem{FMNB:03}
  R.~J.~Fries, B.~M\"uller, C.~Nonaka and S.~A.~Bass,
  \PRL{90,2003,202303}.

\bibitem{Fries:2003kq}
  R.~J.~Fries, B.~M\"uller, C.~Nonaka and S.~A.~Bass,
  \PRC{68,2003,044902}.

\bibitem{Greco:2003xt}
  V.~Greco, C.~M.~Ko and P.~Levai,
  \PRL{90,2003,202302}.


\bibitem{Voloshin:2002wa}
  S.~A.~Voloshin,
  \NPA{715,2003,379c}.

\bibitem{MoVo:03}
  D.~Moln$\acute{\rm a}$r and S.~A.~Voloshin,
  \PRL{91,2003,092301}.

\bibitem{Kolb:2003dz}
  P.~F.~Kolb and U.~W.~Heinz,
  nucl-th/0305084.

\bibitem{Huovinen:2003fa}
  P.~Huovinen,
  nucl-th/0305064.

\bibitem{Hirano:2002hv}
  T.~Hirano and K.~Tsuda,
  \NPA{715,2003,821}.

\bibitem{Nonaka:2006yn}
  C.~Nonaka and S.~A.~Bass,
  nucl-th/0607018.

\bibitem{Adcox:2002pe}
  K.~Adcox {\it et al.}  [PHENIX Collaboration],
  \PLB{561,2003,82}.

\bibitem{Adler:2002xw}
  C.~Adler {\it et al.}  [STAR Collaboration],
  \PRL{89,2002,202301}.

\bibitem{Bjorken:1982tu}
  J.~D.~Bjorken,
  FERMILAB-PUB-82-059-THY (unpublished).
  
\bibitem{Thoma:1990fm}
  M.~H.~Thoma and M.~Gyulassy,
  \NPB{351,1991,491}.

\bibitem{Wang:1991xy}
  X.~N.~Wang and M.~Gyulassy,
  \PRL{68,1992,1480}.

\bibitem{Gyulassy:1993hr}
  M.~Gyulassy and X.~Wang,
  \NPB{420,1994,583}.

\bibitem{Baier:1996kr}
  R.~Baier, Y.~L.~Dokshitzer, A.~H.~Mueller, S.~Peigne and D.~Schiff,
  \NPB{483,1997,291}.

\bibitem{Zakharov:1997uu}
  B.~G.~Zakharov,
  \JL{JETP Lett.,65,1997,615}.

\bibitem{Gyulassy:2000fs}
  M.~Gyulassy, P.~Levai and I.~Vitev,
  \PRL{85,2000,5535}.


\bibitem{Wiedemann:2000za}
  U.~A.~Wiedemann,
  \NPB{588,2000,303}.

\bibitem{Baier:2001yt}
  R.~Baier, Y.~L.~Dokshitzer, A.~H.~Mueller and D.~Schiff,
  \JHEP{0109,2001,033}.

\bibitem{Peshier:1999ww}
  A.~Peshier, B.~K\"ampfer and G.~Soff,
  \PRC{61,2000,045203}.

\bibitem{Schneider:2001nf}
  R.~A.~Schneider and W.~Weise,
  \PRC{64,2001,055201}.

\bibitem{Thaler:2003uz}
  M.~A.~Thaler, R.~A.~Schneider and W.~Weise,
  \PRC{69,2004,035210}.

\bibitem{Bluhm:2004xn}
  M.~Bluhm, B.~K\"ampfer and G.~Soff,
  \PLB{620,2005,131}.

\bibitem{Blaizot:2003iq}
  J.~P.~Blaizot, E.~Iancu and A.~Rebhan,
  \PRD{68,2003,025011}.

\bibitem{Andersen:2003zk}
  J.~O.~Andersen, E.~Petitgirard and M.~Strickland,
  \PRD{70,2004,045001}.

\bibitem{Gavai:2005sd}
  R.~V.~Gavai and S.~Gupta,
  \PRD{72,2005,054006}.

\bibitem{Koch:2005vg}
  V.~Koch, A.~Majumder and J.~Randrup,
  \PRL{95,2005,182301}.

\bibitem{Majumder:2006nq}
  A.~Majumder and B.~M\"uller,
  nucl-th/0605079.

\bibitem{Liao:2005pa}
  J.~Liao and E.~V.~Shuryak,
  \PRD{73,2006,014509}.

\bibitem{Molnar:2004yh}
  D.~Molnar and P.~Huovinen,
  \PRL{94,2005,012302}.

\bibitem{Teaney:2006nc}
  D.~Teaney,
  hep-ph/0602044.

\bibitem{Kovtun:2006pf}
  P.~Kovtun and A.~Starinets,
  \PRL{96,2006,131601}.

\bibitem{Gyulassy:2004zy}
  M.~Gyulassy and L.~McLerran,
  \NPA{750,2005,30}.

\bibitem{Gubser:1998nz}
  S.~S.~Gubser, I.~R.~Klebanov and A.~A.~Tseytlin,
  \NPB{534,1998,202}.

\bibitem{Policastro:2001yc}
  G.~Policastro, D.~T.~Son and A.~O.~Starinets,
  \PRL{87,2001,081601}.

\bibitem{Buchel:2004di}
  A.~Buchel, J.~T.~Liu and A.~O.~Starinets,
  \NPB{707,2005,56}.

\bibitem{Liu:2006ug}
  H.~Liu, K.~Rajagopal and U.~A.~Wiedemann,
  hep-ph/0605178.

\bibitem{Casalderrey:2006rq}
  J.~Casalderrey-Solana and D.~Teaney,
  hep-ph/0605199.

\bibitem{Herzog:2006gh}
  C.~P.~Herzog, A.~Karch, P.~Kovtun, C.~Kozcaz and L.~G.~Yaffe,
  hep-th/0605158.

\bibitem{Herzog:2006se}
  C.~P.~Herzog,
  hep-th/0605191.

\bibitem{Shuryak:2004tx}
  E.~V.~Shuryak and I.~Zahed,
  \PRD{70,2004,054507}.

\bibitem{Gelman:2006xw}
  B.~A.~Gelman, E.~V.~Shuryak and I.~Zahed,
  nucl-th/0601029.

\bibitem{Huot:2006ys}
  S.~C.~Huot, S.~Jeon and G.~D.~Moore,
  hep-ph/0608062.

\bibitem{Weibel:1959}
  E.~S.~Weibel,
  \PRL{2,1959,83}.

\bibitem{Mrowczynski:1996vh}
  S.~Mrowczynski,
  \PLB{393,1997,26}.

\bibitem{Arnold:2003rq}
  P.~Arnold, J.~Lenaghan and G.~D.~Moore,
  \JHEP{0308,2003,002}.

\bibitem{Randrup:2003cw}
  J.~Randrup and S.~Mrowczynski,
  \PRC{68,2003,034909}.

\bibitem{Arnold:2004ti}
  P.~Arnold, J.~Lenaghan, G.~D.~Moore and L.~G.~Yaffe,
  \PRL{94,2005,072302}.

\bibitem{Mueller:2005un}
  A.~H.~Mueller, A.~I.~Shoshi and S.~M.~H.~Wong,
  \PLB{632,2006,257}.

\bibitem{Arnold:2004ih}
  P.~Arnold and J.~Lenaghan,
  \PRD{70,2004,114007}.

\bibitem{Mrowczynski:2005ki}
  S.~Mrowczynski,
  \JL{Acta Phys.\ Polon.\ B,37,2006,427}.

\bibitem{Schenke:2006xu}
  B.~Schenke, M.~Strickland, C.~Greiner and M.~H.~Thoma,
  hep-ph/0603029.

\bibitem{Dupree:1966}
  T.~H.~Dupree,
  \JL{Phys.\ Fluids,9,1966,1773}.

\bibitem{Dupree:1968}
  T.~H.~Dupree,
  \JL{Phys.\ Fluids,11,1968,2680}.

\bibitem{Malone:1975}
  R.~C.~Malone, R.~L.~McCrory, and R.~L.~Morse,
  \PRL{34,1975,721}.

\bibitem{Okada:1978}
  T.~Okada, T.~Yabe, and K.~Niu,
  \JL{J.\ Plasma Phys.,20,1978,405}.

\bibitem{Abe:1980a}
  T.~Abe and K.~Niu,
  \JPSJ{49,1980,717}.

\bibitem{Abe:1980b}
  T.~Abe and K.~Niu,
  \JPSJ{49,1980,725}.

\bibitem{Okada:1980}
  T.~Okada and K.~Niu,
  \JL{J.\ Plasma Phys.,23,1980,423}.

\bibitem{Danielewicz:1984ww}
  P.~Danielewicz and M.~Gyulassy,
  \PRD{31,1985,53}.

\bibitem{Heinz:1983nx}
  U.~W.~Heinz,
  \PRL{51,1983,351}.

\bibitem{Heinz:1984yq}
  U.~W.~Heinz,
  \ANN{161,1985,48}.

\bibitem{Heinz:1985qe}
  U.~W.~Heinz,
  \ANN{168,1986,148}.

\bibitem{Rebhan:2004ur}
  A.~Rebhan, P.~Romatschke and M.~Strickland,
  \PRL{94,2005,102303}.

\bibitem{Landau10}
  E.~M.~Lifshitz and L.~P.~Pitaevskii,
  \textit{Physical Kinetics},
  (Pergamon Press, Oxford-New York, 1981).

\bibitem{Israel:1976tn}
  W.~Israel,
  \ANN{100,1976,310}.

\bibitem{Israel:1979wp}
  W.~Israel and J.~M.~Stewart,
  \ANN{118,1979,341}.

\bibitem{Chen:2006ig}
  J.~W.~Chen and E.~Nakano,
  hep-ph/0604138.

\bibitem{Bjorken:1982qr}
  J.~D.~Bjorken,
  \PRD{27,1983,140}.

\bibitem{Heiselberg:1995sh}
  H.~Heiselberg and X.~N.~Wang,
  \PRC{53,1996,1892}.

\bibitem{Selikhov:1993ns}
  A.~Selikhov and M.~Gyulassy,
  \PLB{316,1993,373}.

\bibitem{Arnold:1998cy}
  P.~Arnold, D.~T.~Son and L.~G.~Yaffe,
  \PRD{59,1999,105020}.

\bibitem{deGroot}
  S.~R.~deGroot, W.~A.~van Leeuwen, and Ch.~G.~van Weert
  \textit{Relativistic Kinetic Theory},
  (North-Holland, Amsterdam-New York-Oxford, 1980).

\bibitem{Fermi:1949ee}
  E.~Fermi,
  \PR{75,1949,1169}.

\bibitem{Baier:2000mf}
  R.~Baier, D.~Schiff and B.~G.~Zakharov,
  \JL{Ann.\ Rev.\ Nucl.\ Part.\ Sci.,50,2000,37}.

\bibitem{Gelis:2005pb}
  F.~Gelis, K.~Kajantie and T.~Lappi,
  \PRL{96,2006,032304}.

\bibitem{Hufner:1989fn}
  J.~H\"ufner, B.~Povh and S.~Gardner,
  \PLB{238,1990,103}.

\bibitem{Wang:2004kf}
  F.~Wang  [STAR Collaboration],
  \JL{J.\ Phys.\ G,30,2004,S1299}.

\bibitem{Rebhan:2005re}
  A.~Rebhan, P.~Romatschke and M.~Strickland,
  \JHEP{0509,2005,041}.

\bibitem{Arnold:2005vb}
  P.~Arnold, G.~D.~Moore and L.~G.~Yaffe,
  \PRD{72,2005,054003}.

\end{thebibliography}
\end{document}